\documentclass[11pt,twoside]{book} 
\usepackage{asp2008s}
\usepackage{times}
\usepackage{lscape}
\usepackage{epsf}

\usepackage{graphicx}
\usepackage{amssymb}
\usepackage{longtable}
\usepackage[figuresright]{rotating}
\usepackage{multirow}

\newcommand{\Lsun}{L$_{\odot}$}

\newcommand{\simless}{\mathbin{\lower 3pt\hbox {$\rlap{\raise 5pt\hbox{$\char'074$}}\mathchar"7218$}}}

\mainmatter

\newlength{\deftabcolsep}
\begin{document}

\title{The Serpens Molecular Cloud}
\author{C. Eiroa}
\affil{Dpto. F\'\i sica Te\'orica, Universidad Aut\'onoma de Madrid, 28049
Madrid, Spain}
\author{A. A. Djupvik}
\affil{Nordic Optical Telescope, Apdo 474, 38700 Santa Cruz de La Palma, Spain}
\author{M. M. Casali}
\affil{European Southern Observatory, Karl-Schwarzschild-Strasse 2, 85748
Garching, Germany}

\begin{abstract}
The  Serpens cloud  has received  considerable attention  in  the last
years, in particular the small  region known as the Serpens cloud core
where a plethora  of star formation related phenomena  are found. This
review summarizes  our current  observational knowledge of  the cloud,
with emphasis on the core. Recent results are converging to a distance
for  the cloud  of  $\sim 230  \pm 20$  pc,  an issue  which has  been
controversial over the  years. We present the gas  and dust properties
of  the  cloud core  and  describe  its  structure and  appearance  at
different wavelengths. The core contains a dense, very young, low mass
stellar cluster with more than 300 objects in all evolutionary phases,
from collapsing  gaseous condensations to pre-main  sequence stars. We
describe  the  behaviour and  spatial  distribution  of the  different
stellar  populations (mm  cores, Classes  0,  I and  II sources).  The
spatial concentration and the fraction number of Class 0/Class I/Class
II  sources is considerably  larger in  the Serpens  core than  in any
other  low  mass star  formation  region,  e.g.  Taurus, Ophiuchus  or
Chamaeleon, as also stated  in different works. Appropriate references
for coordinates and fluxes of  all Serpens objects are given. However,
we provide  for the first time  a unified list of  all near-IR sources
which have  up to now been  identified as members of  the Serpens core
cluster; this list includes some  members identified in this review. A
cross-reference  table of  the near-IR  objects with  optical, mid-IR,
submillimeter, radio  continuum and X-ray  surces is also  provided. A
simple analysis has allowed us to identify a sample of $\sim 60$ brown
dwarf  candidates among  the 252  near-IR objects;  some of  them show
near-IR excesses and, therefore,  they constitute an attractive sample
to study very young substellar  objects. The review also refers to the
outflows associated with the young  sources. A section is dedicated to
the  relatively small  amount  of works  carried  out towards  Serpens
regions outside the core. In particular, we refer to ISO and to recent
Spitzer  data.  These  results  reveal  new  centers  of  active  star
formation in the Serpens cloud and the presence of new young clusters,
which deserve  follow-up observations  and studies to  determine their
characteristics  and  nature in  detail.  Finally,  we  give a  short,
non-exhaustive list of individually interesting Serpens objects.

\end{abstract}

\section{Introduction}
\label{introduction}
\begin{figure}[!ht]
\begin{center}
\scalebox{0.60}{
\includegraphics{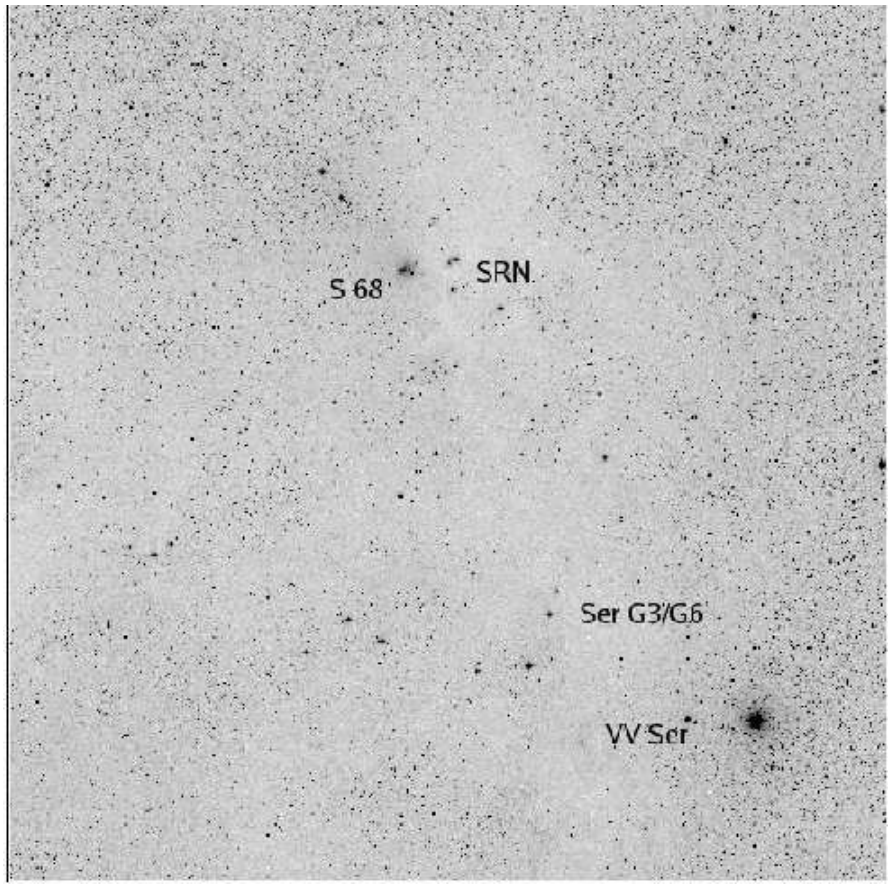}
}
\caption{The  Serpens molecular  cloud  as seen  on  the DSS  $R$-band
plates. Large  scale irregular dark  structures are clearly  seen. The
reflection nebulosities  S 68  and SRN are  indicated, as well  as the
position  of VV  Ser  and of  the  H$\alpha$ emission  line stars  Ser
G3/G6.   Field   size  is   $2\deg   \times   2\deg$.  Field   centre:
$\alpha_{2000}    =   18^h$    30$^m$,    $\delta_{2000}   =    0\deg$
$50\arcmin$. North  to the top, East  to the left. The  image has been
downloaded    from     the    Canadian    Astronomy     Data    Center
(http://cadcwww.dao.nrc.ca/dss).}
\label{Serpens}
\end{center}
\end{figure}

A considerable amount  of work has been dedicated  to the region known
as the  Serpens dark cloud  (Galactic coordinates $l^{II} =  32\deg, ~
b^{II} = 5\deg$)  since it was recognized by Strom et  al. (1974) as a
site  of active  star  formation. The  cloud  extends several  degrees
around the  young variable  star VV  Ser and forms  part of  the large
local  dark cloud  complex  called  the Aquila  Rift,  which has  been
extensively mapped  in several molecular  line surveys (e.g.   Dame \&
Thaddeus 1985,  Dame et al.  1987,  2001, see the chapter  by Prato et
al.). A  large scale extinction  map has been presented  by Cambr\'esy
(1999) and Dobashi et  al.  (2005). Serpens is seen  on optical images
as  a  large area  irregularly  obscured  by  large amounts  of  dust;
Fig. \ref{Serpens} reproduces the DSS  $R$ band photograph of a region
$2\deg  \times 2\deg$  in  size. Several  reflection nebulosities  are
distinguished, the most  prominent of which are S  68 (Sharpless 1959,
see  also Dorschner  \& G\"urtler  1963,  van den  Bergh 1966,  Bernes
1977), illuminated  by HD 170634,  and a very red  bipolar, reflection
nebulosity,  usually referred to  as the  Serpens object,  the Serpens
nebula, or the Serpens reflection nebulosity (SRN), illuminated by the
pre-main  sequence star  SVS 2  (Strom et  al. 1974,  1976,  Worden \&
Grasdalen 1974, King  et al.  1983, Warren-Smith et  al. 1987, G\'omez
de Castro et al. 1988). In this work, we will refer to that nebulosity
as SRN.

Relatively little  work has been  devoted to studying the  large scale
properties  of  the Serpens  cloud,  having  been mainly  concentrated
around SRN  and VV  Ser, which  are located towards  the North  of the
large  extinction map  analysed by  Cambr\'esy (1999).  Cohen  \& Kuhi
(1979) and Chavarr\'\i  a et al.  (1988) conducted optical  studies of
the region  and identified several  H$\alpha$ emission line  stars and
some B,  A and  later spectral type  stars associated  with reflection
nebulae. Loren et al. (1979) carried out  a 2.6 mm line CO map over an
area of  $\sim$ 300  square arc minutes  and detected a  dense H$_2$CO
core of  $\sim ~3\farcm3$  in radius  centred in SRN.  IRAS maps  of a
$\sim ~3\deg  \times 3\deg$  region around the  cloud core  show large
scale  extended   far-IR  emission  and   several  point-like  sources
associated   with  some   visible  and   invisible  stars   (Zhang  et
al. 1988b). The cloud core with several point-like sources embedded in
extended emission shows  up very prominently in all  IRAS bands (Zhang
et  al. 1988a).  More  recently  Schnee et  al.  (2005) have  analysed
recalibrated  IRAS data  to estimate  dust column  densities  and dust
temperatures.   Assuming a  distance of  700 pc  to Serpens,  Zhang et
al. (1988b) estimate a total  far-IR luminosity from the whole Serpens
cloud of  $\sim$ 3200  L$_\odot$ and the  dust mass inferred  from the
extended  emission is  $\sim$ 290  M$_\odot$. The  actual  distance to
Serpens is most likely significantly smaller than the value assumed by
Zhang et  al. and,  therefore, the luminosity  and dust  emitting mass
should  be corrected.  In fact,  the distance  to Serpens  has  been a
controversial  point over  the years,  with suggested  figures  in the
range  from $\sim$  200 pc  to $\sim$  700 pc.  At present,  a smaller
distance value is commonly accepted. We analyse this issue in the next
section.

After  the discovery  of the  cloud core,  observational  efforts have
concentrated in that area. The  reason is the extreme richness in star
formation  activity found  in the  core, with  all  relevant phenomena
taking place  in a small region  of just few arc  minutes in diameter:
pre-protostellar     gaseous    condensations;     pre-stellar    dust
condensations;  Class  0, Class  I  and  Class  II objects;  different
evidences of  infalling gas; disks; molecular and  atomic outflows and
jets;  clustering  of  young   stellar  objects  (YSOs)  in  different
evolutionary stages, etc. The  Serpens core indeed represents a unique
laboratory  for studies  of  star formation  processes and  observable
phenomena, and the inter-relation between them.

A first review  article of the most relevant  observational results up
to  $\sim  1991$  was  written   by  us  (Eiroa  1991).  In  this  new
contribution we concentrate mainly on results achieved during the last
15  years. We  review  observations  related to  the  Serpens core  in
particular, since this part of the cloud has continued to be the focus
of  the  large observational  effort  carried  out  by many  different
research groups.  Nonetheless, a section  is also dedicated  to recent
work carried  out in Serpens areas  outside the core.  The new results
confirm and emphasize the role of Serpens as a key region for detailed
studies  of  star  formation  and  the understanding  of  all  related
phenomena.

\section{The Distance of Serpens}

The  distance to Serpens  has been  a matter  of controversy  over the
years.  Until 1996  the distance  to the  Serpens molecular  cloud was
estimated  from   a  handful  of  stars   associated  with  reflection
nebulae. Mainly through disagreements  on spectral type and luminosity
classes,  the different distances  obtained were  the subject  of much
debate (see Strai\u{z}ys et al.  1996 for details). Racine (1968) gave
a first distance estimate (440 pc) based on the star HD170634 which he
classified as a B7 V star. Strom et al. (1974) found the same distance
to the same star independently from their IR data classifying it as an
A0  (V:).   Chavarria-K  et  al. (1988)  obtained  spectroscopy,  {\em
uvby$\beta$} and JHKLM  photometry for this one (A1  V) and three more
stars: SAO 123590 (B4 V), SAO  123595 (B3 V), and SAO 123661 (B3), and
obtained a distance of 245 $\pm$  30 pc. Zhang et al. (1988), however,
estimated 700  pc (with a large  scatter) for three of  the above four
stars,  having assigned  luminosity class  III (citing  an unpublished
reference) and found support from kinematics (CO and NH$_3$ lines). De
Lara  \& Chavarria (1989)  rechecked their  spectra and  confirmed the
main sequence nature of the  above sources, and moreover estimated d =
296 $\pm$ 34  pc based on 7 stars. Excluding  HD170634 (claiming it to
be a shell star) and adding two other reflection nebulae stars [CDF88]
1 and [CDF88] 7  de Lara et al. (1991) arrived at d  = 311 $\pm$ 38 pc
and a mean  total to selective extinction ratio $R  = A_V/E(B-V) = 3.3
\pm 0.3$.  This estimate was in  use until Strai\u{z}ys  et al. (1996)
revised  the  distance using  a  completely  independent method.  They
surveyed  photometrically most  stars down  to magnitude  13 in  a 6.5
square  degrees region  in  the Vilnius  photometric  system of  seven
narrow bands ($UPXYZVS$). Spectral  types and absolute magnitudes were
found for  105 stars -  of which 18  belong to the cloud.  This sample
included the earlier used stars, and it was noted that both SAO 123590
and SAO  123661 are close  binaries.  Applying the total  to selective
extinction ratio found  by de Lara et al.  (1991) for  the stars in or
behind  the molecular  cloud they  found a  distance of  259  $\pm$ 37
pc. This estimate was also supported  by Festin (1998) who made use of
the cloud as  a screen when searching for foreground  dwarfs in a 0.15
square  degree  area  about  4  degrees  to the  South  of  the  Cloud
Core. Strai\u{z}ys et al. (2003)  then investigated 473 stars across a
larger area  (5 $\times$ 10 degrees)  and found the front  edge of the
clouds in  this area to be  at 225 $\pm$  55 pc and the  cloud complex
possibly 80 pc deep. These are  believed to be part of the Aquila Rift
large dark  cloud complex, located at  a distance of 200  $\pm$ 100 pc
(Dame \&  Thaddeus 1985). Using  2MASS data Knude (2005)  recently has
estimated extinction and  distances for 754 stars in  the same area as
that used by Straizys et al. (1996) and find a distance to the Serpens
cloud  of 210 $\pm$  12 pc,  estimating the  extinction jump  by using
several physical indicators, among  these the A$_V$ vs. $\sigma_{A_V}$
variation seen in MHD simulations.  Thus, all recent and, likely, more
reliable results are converging to a distance $\sim 230 \pm 20$ pc for
the Serpens cloud.

\section{An Overall Description of the Serpens Core}

\begin{figure}[p]
\begin{center}
\scalebox{0.45}{
\includegraphics{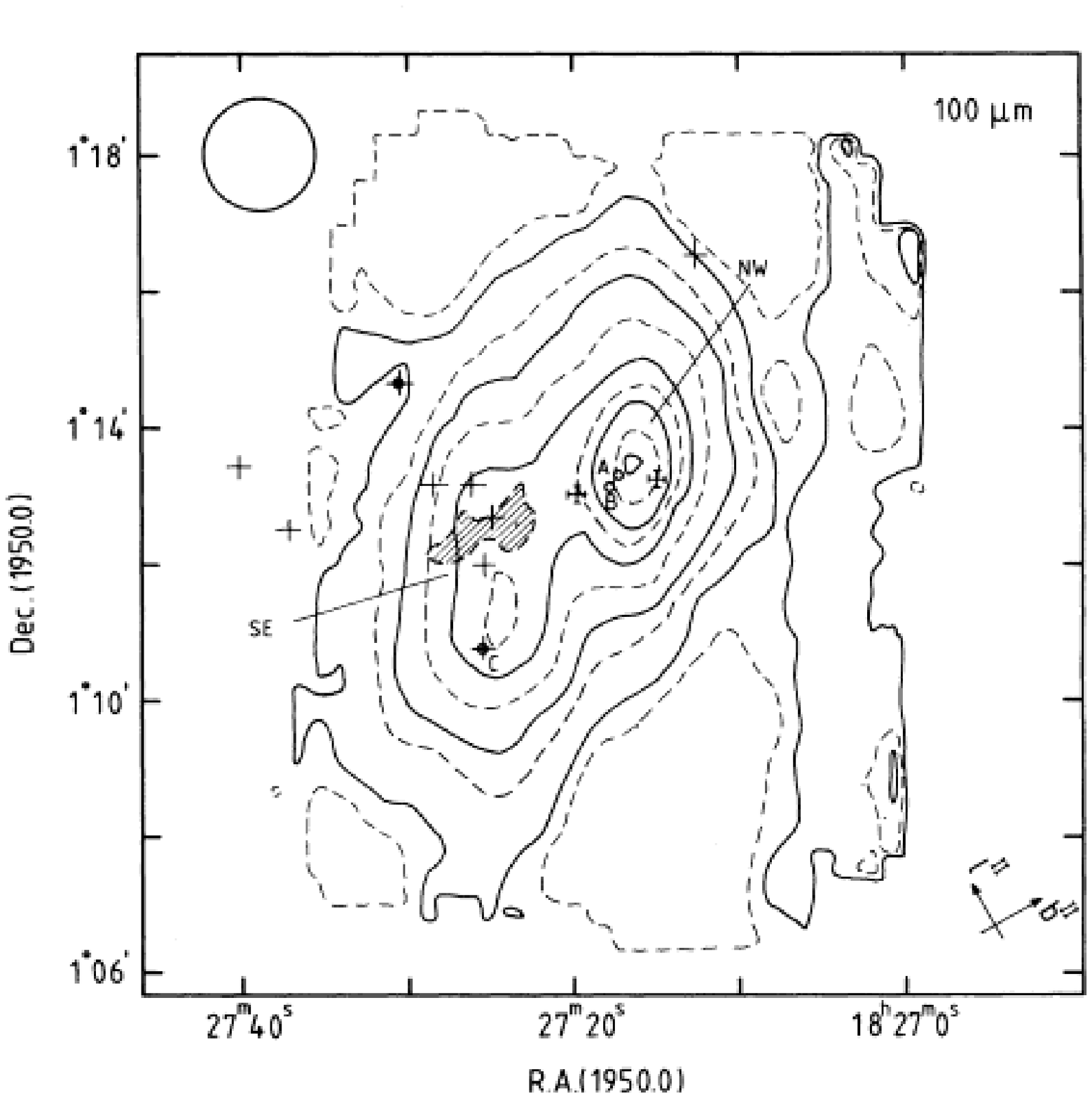}
}
\caption{IRAS-CPC 100 $\mu$m map of the Serpens core (taken from Zhang
et  al.,  1988a).  The  dashed  region  approximately  represents  the
position of SRN.}
\label{IRAS}
\end{center}
\vspace{10mm}
\begin{center}
\scalebox{0.42}{
\includegraphics{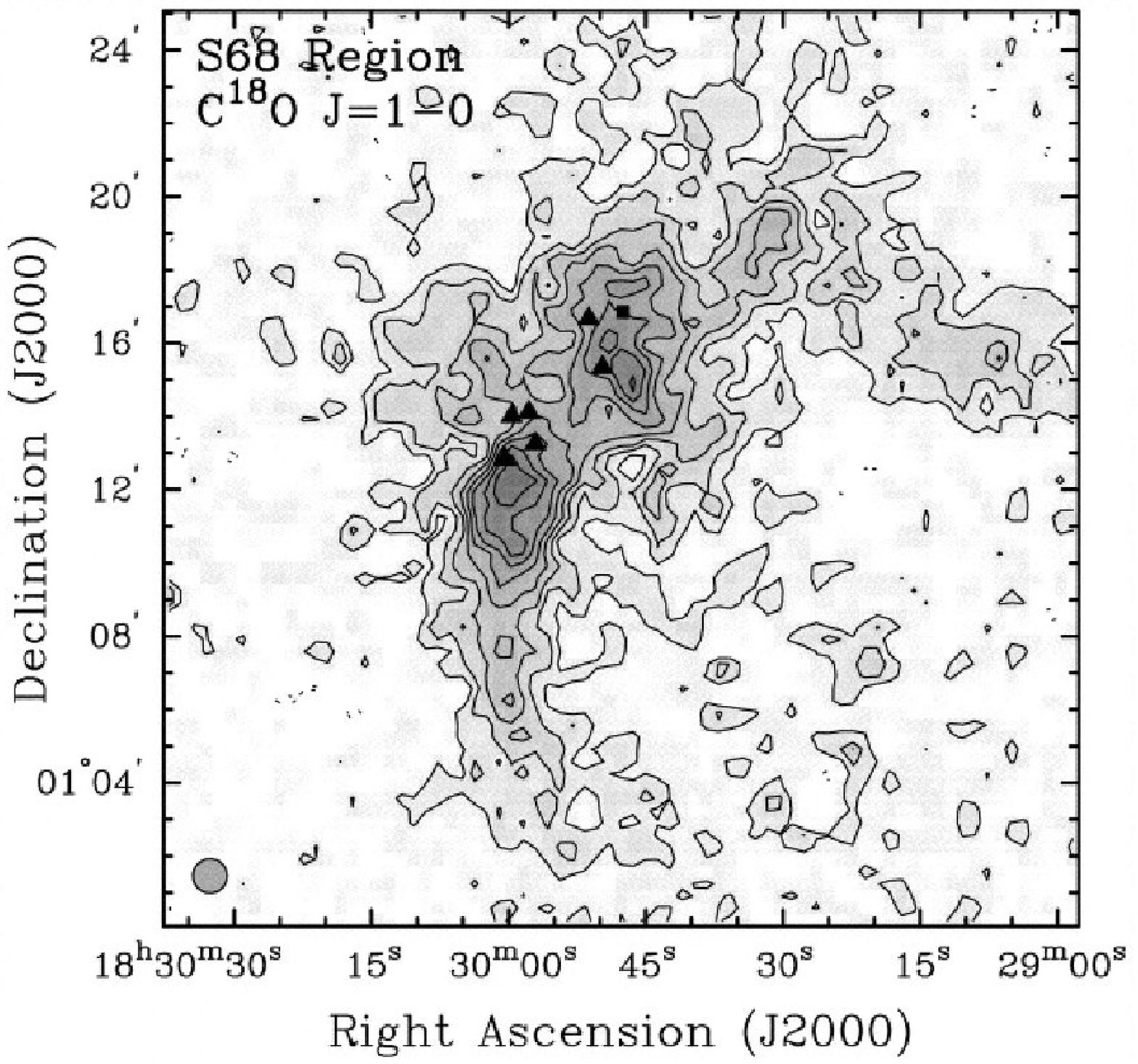}
}
\caption{Integrated  C$^{18}$O J=1-0  emission from  the  Serpens core
(taken from McMullin et al., 2000). Black symbols represent some young
embedded objects.}
\label{CO}
\end{center}
\end{figure}

The early observations of Loren et al. (1979) found a nearly circular,
high density H$_2$CO core approximately centred in SRN, while maps of
the  far-IR  emission,  NH$_3$  and  other  molecules  showed  a  more
elongated  structure extending  in  a North-West/South-East  direction
with  two broad peaks:  a Southeastern  (SE) clump  with a  broad peak
close  to SRN, extending  towards the  South from  this object,  and a
second  Northwestern (NW)  clump peaking  at a  position close  to the
source FIRS 1, the most  prominent object embedded in the Serpens core
(e.g.  Ho \&  Barrett 1980,  Little et  al. 1980,  Nordh et  al. 1982,
Ungerechts \& G\"usten 1984, Harvey et al. 1984, Takano 1986, Zhang et
al.   1988a, Torrelles et  al. 1989).  Estimated 60/100  $\mu$m colour
temperatures  in  the range  $\sim$  20  - 30  K  are  similar to  the
kinematic gas temperatures found by  Loren et al. (1979). The division
of the internal structure of the  gas and dust into two main subclouds
or clumps  has been confirmed by recent  studies, including HIRES-IRAS
images (Hurt  \& Barsony, 1996), submm/mm continuum  maps (e.g. Casali
et al. 1993, Davis et al. 1999, Kaas et al. 2004), and maps of several
molecular lines  coming from different isotopical  species and tracing
gas  with different  column densities,  like  CO,$^{13}$CO, C$^{18}$O,
H$_2$CO,  HCO$^+$,  N$_2$H$^+$,  CS,   HCN,  etc.  (e.g.  McMullin  et
al.    1994,   2000,   White    et   al.    1995,   Testi    et   al.,
2000). Fig. \ref{IRAS}  and \ref{CO} show the IRAS  and C$^{18}$O maps
of the Serpens core respectively  (Zhang et al. 1988a, and McMullin et
al. 2000), where  the extended core emission and the  two broad SE and
NW clumps  of dust and gas show  up very clearly. Davis  et al. (1999)
find some cavity-like submm structures (see Fig. \ref{Davis_smm}); in
particular, the  most remarkable  one is found  between the NW  and SE
clumps  (see  also  Fig.  11  in  Kaas  et  al,  2004).  In  addition,
small-scale gas  and dust structures within each  individual clump are
seen in high resolution maps  (e.g. Fig. \ref{CO}).  The complexity of
the cloud  core reflects density and temperature  enhancements and the
effects of the embedded  stellar/protostellar population; Davis et al.
(1999) suggest  that the  cavity-like submm  structures may  have been
shaped by outflows from young objects in the core. Both NW and SE core
clumps  behave kinematically different  (Testi et  al. 2000);  in this
respect, it  is interesting  to note that  the characteristics  of the
embedded  YSO  population  within  each  individual  clump  show  some
differences, at least when the  evolution of the sources is considered
(see Sect. 5).

While  there is  an overall  agreement  between the  results found  by
different  groups concerning the  structure, temperature,  density and
gas   kinematics,  a   detailed  comparison   shows  some   points  of
disagreement.  Those could be due  to differences in beam sizes and/or
the  parameters and  methods used  to analyze  the data.  It  would be
interesting  to  clarify  these  discrepancies  by  carrying  out  new
specific observations and  analysis. Many molecules show self-absorbed
line profiles in  different core positions (see Loren  \& Wootten 1980
and Loren et al. 1981, for the earliest works), in particular close to
embedded  protostellar   objects,  and  their   kinematical  behaviour
suggests infall motions; furthermore, the  SE and NW clumps seem to be
undergoing  global,  large-scale  ($\sim$  0.2 pc)  infalling  motions
(Williams \& Myers 2000, Olmi \& Testi 2002). The latter authors point
out that  the data are  in qualitative agreement with  the theoretical
expectation of a slow contraction  followed by rapid star formation in
localized high density regions. A  velocity gradient from East to West
of $\sim$ 1  - 2 km s$^{-1}$ pc$^{-1}$ is observed  in the most recent
C$^{18}$O maps (McMullin et al.  2000, Olmi \& Testi 2002), supporting
the early  NH$_3$ velocity gradient claimed by  Ungerechts \& G\"usten
(1984); in  apparent  contradiction  with previous  C$^{18}$O  results
(White et al. 1995), which did not find any evidence for such velocity
gradient.   Depending   on  the   molecules   and  transition   lines,
temperatures  are found  in the  range from  $\sim$ 15  K in  the more
diffuse parts of the core, up to $\sim$ 50 K in condensations close to
embedded young  objects. The mean average temperature  across the core
is $\sim$  25 - 30 K  (McMullin et al.  2000).  These gas temperatures
are similar  to dust temperatures found  across the cloud  core and in
the  dust surrounding  the embedded  YSOs  (e.g. Zhang  et al.  1988a,
Casali et  al. 1993, Hurt \&  Barsony 1996, Larsson et  al. 2000). Gas
particle densities n(H$_2$) are found  in the range from $\sim$ 10$^4$
cm$^{-3}$ to values larger than 10$^6$ cm$^{-3}$. Mass estimates based
on the C$^{18}$O  (J = 1 -- 0) transition line  give values of $\sim$
250 - 300 M$_\odot$ (McMullin et al. 2000, Olmi \& Testi, 2002), while
the C$^{18}$O (J = 2 --  1) line gives $\sim$ 1450 M$_\odot$ (White et
al. 1995). McMullin et al. (2000)  argue that the estimate made on the
J =  1 -- 0 line is  better than that based  on J = 2  -- 1 transition
because the first is less sensitive to density and has a factor of 2-3
less opacity. Even allowing for larger uncertainties (but unrelated to
the cloud distance), the difference  seems to be significant and it is
an open issue to  be resolved. We remark that the cloud  mass is a key
parameter in  evaluating the star formation efficiency  and the global
star formation properties of the cloud.

The near-IR  study of  the dust properties  by Eiroa \&  Hodapp (1989)
showed  that  water  ice  absorption  is  widespread  in  the  Serpens
molecular  cloud.  They also  detected  solid  CO absorption,  further
studied by Chiar et al. (1994),  finding an unusually high ratio of CO
to H$_2$O  ice absorption.  Ice features in  the near- and mid-IR  using
Spitzer  mid-IR spectroscopy and ground-based facilities have
recently been  studied  by  Knez et  al.  (2005), Pontoppidan et al.
(2008), Boogert et al. (2008), and \"Oberg et al. (2008). The structure
of the  CO ice absorption has been studied
by  Pontoppidan et al.  (2003b). Pontoppidan  et al.  (2003a) detected
methanol in  the wings  of the  3.1 ~$\mu$m water  ice feature  in the
lines of  sight to  SVS 4-2 and  SVS 4-9,  the first detection  of the
molecule toward low-luminosity young  stars. Pontoppidan et al. (2004)
subsequently  extended the $L$-band  observations to  the whole  SVS 4
stellar  complex  (Eiroa \&  Casali  1989).  Big  jumps in  calculated
extinction and water ice absorption  were found from source to source,
even over  projected distances of  only 10 arcseconds.  Thus,  some of
the  sources probably lie  behind and  some in  front of  an obscuring
cloud, as well as being embedded in  the envelope of the Class 0 SMM 4
object (Casali et al. 1993). The water ice optical depth-to-extinction
ratio was used  to infer abundances of $9  \times 10^{-5}$ relative to
$H_2$, corresponding to water ice  mantle volumes around 40\% of those
in Taurus. In addition, it  generally seems likely that methanol, with
an  abundance  around  20\%  of  that  of  water,  is  formed  through
grain-surface chemistry in pre-stellar low density conditions and cold
temperatures  ($T <  20K$), rather  than in  the gas  phase.   But the
variation (detection)  in abundance from source to  source is puzzling
and  may   indicate  that  the  methanol  formation   is  a  transient
phenomenon.

The  ISOCAM CVF  study  of  Alexander et  al.  (2003) extended  mid-IR
spectroscopy  to 20  low-luminosity  sources in  Serpens.  SVS 2  (the
illuminating  source of  SRN) clearly  shows the  silicate  feature in
emission - see also Kessler-Silacci  et al. (2006) for the 5-34 $\mu$m
Spitzer   spectrum.   12    other   sources   show   deep   absorption
features.  Water ice  absorption at  6  $\mu$m is  strong relative  to
silicate optical depth, compared to  other regions. On the other hand,
CO$_2$  absorption at  15.2  $\mu$m  is weak  in  Serpens compared  to
silicate depth.  The study shows clear differences  in grain chemistry
from region to region, perhaps related to age or other factors. Ciardi
et  al. (2005)  have recently  resolved  the silicate  feature in  the
binary  SVS 20  and  find that  the  mid-IR spectrum  of  each YSO  is
dominated  by  amorphous   silicate  emission  with  some  crystalline
silicates.  Differences  in the dust properties  toward each component
of the binary might be explained by their different luminosities.

\section{The Embedded Population}

\begin{figure}[!ht]
\begin{center}
\scalebox{0.7}{
\includegraphics{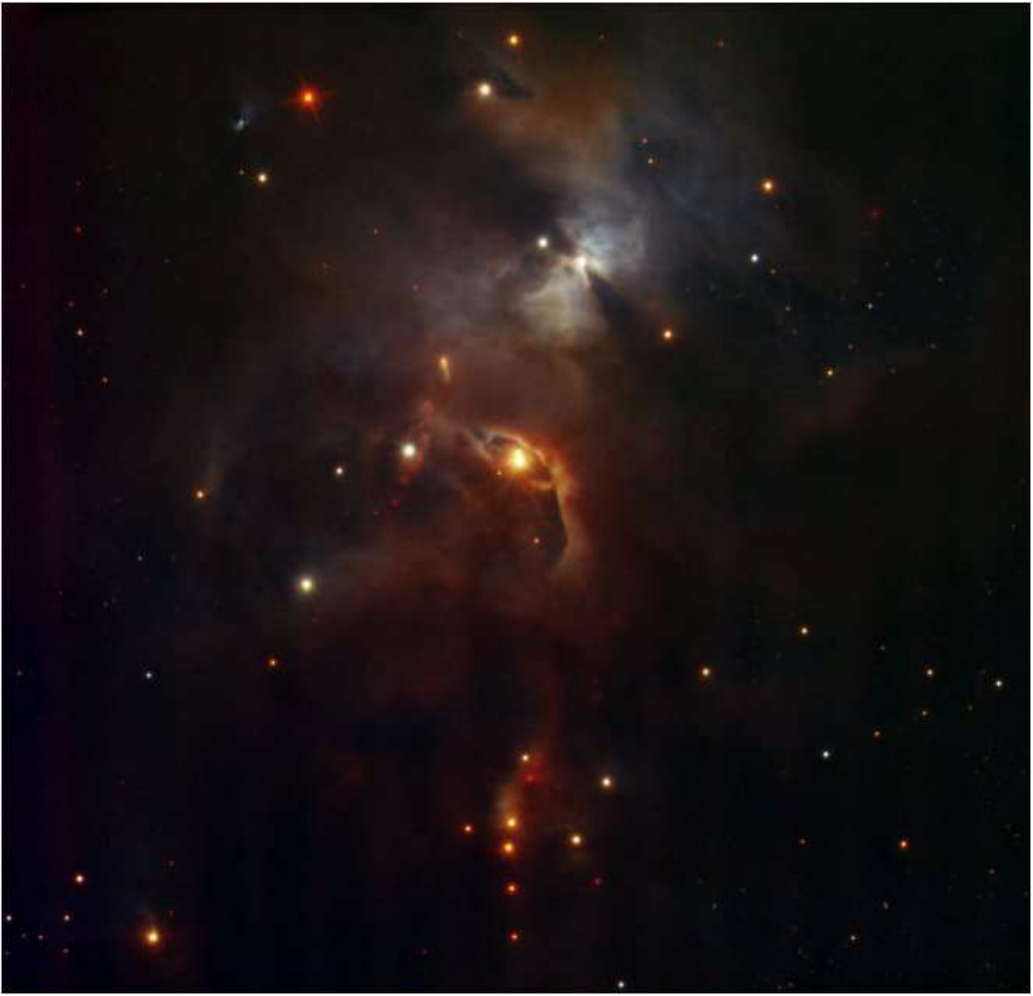}
}
\caption{The Serpens  core and the  embedded near-IR cluster seen  in a
$J$, $H$,  and $K$ composite image  taken by the  ESO's near-IR imager
HAWK-I attached on UT4 of the VLT. The central object surrounded by an
elliptical nebulous ring with spiral-like arms is the young binary SVS
20. At the top, the pre-main sequence  star SVS 2 and its edge-on disk
are clearly  distinguished. The  bottom of the  image shows the  SVS 4
sub-cluster. In  addition, nebulous  objects, $H_2$ emission  knots, a
widespread reflection nebulosity, and many YSOs are clearly seen.}
\label{HAWK}
\end{center}
\end{figure}

Data obtained during the last 15 years in the infrared, submillimeter,
millimeter,  radio  continuum and  X-rays  spectral  domains reveal  a
dense, rich, young,  low mass stellar cluster embedded  in the Serpens
core.   More  than  300  objects  in  all  evolutionary  phases,  from
collapsing  gaseous condensations  to Class  II/III  pre-main sequence
(PMS) stars  coexist within  the $\sim 0.5  - 0.7$ pc  Serpens central
core. Fig.  \ref{HAWK} shows an  impressive colour composite  from
images taken through the near-IR bands $J$, $H$ and $K$ of the Serpens
core.  Individual images have 1 minute exposure each and were recently
obtained during first  light of the HAWK-I near-IR  camera attached on
UT4 of ESO's Very Large Telescope. In the following, we review how the
cluster appears at the different wavelength ranges.

\subsection{The Near- and Mid-IR Cluster: Tracing the Class~I, \\
Flat-spectrum and Class II Object Population}
\label{ClassII_I}

\begin{figure}[!ht]
\begin{center}
\scalebox{0.55}{\rotatebox{-90}{
\includegraphics{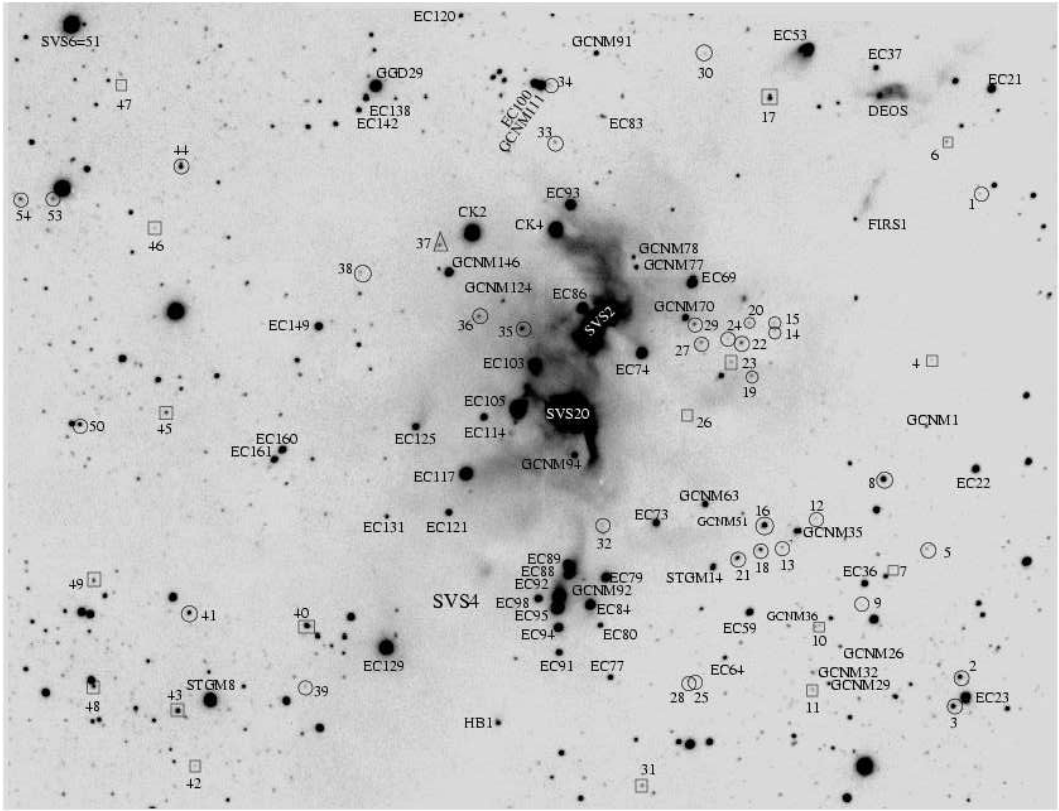}
}}
\caption{K-image of  the Serpens core  (size $\sim 8 \arcmin  \times 6
\arcmin$).  Some of the  embedded objects are identified. Figure taken
from Kaas (1999).}
\label{K-image}
\end{center}
\end{figure}

\begin{figure}[!ht]
\begin{center}
\scalebox{0.40}{\rotatebox{-90}{
\includegraphics{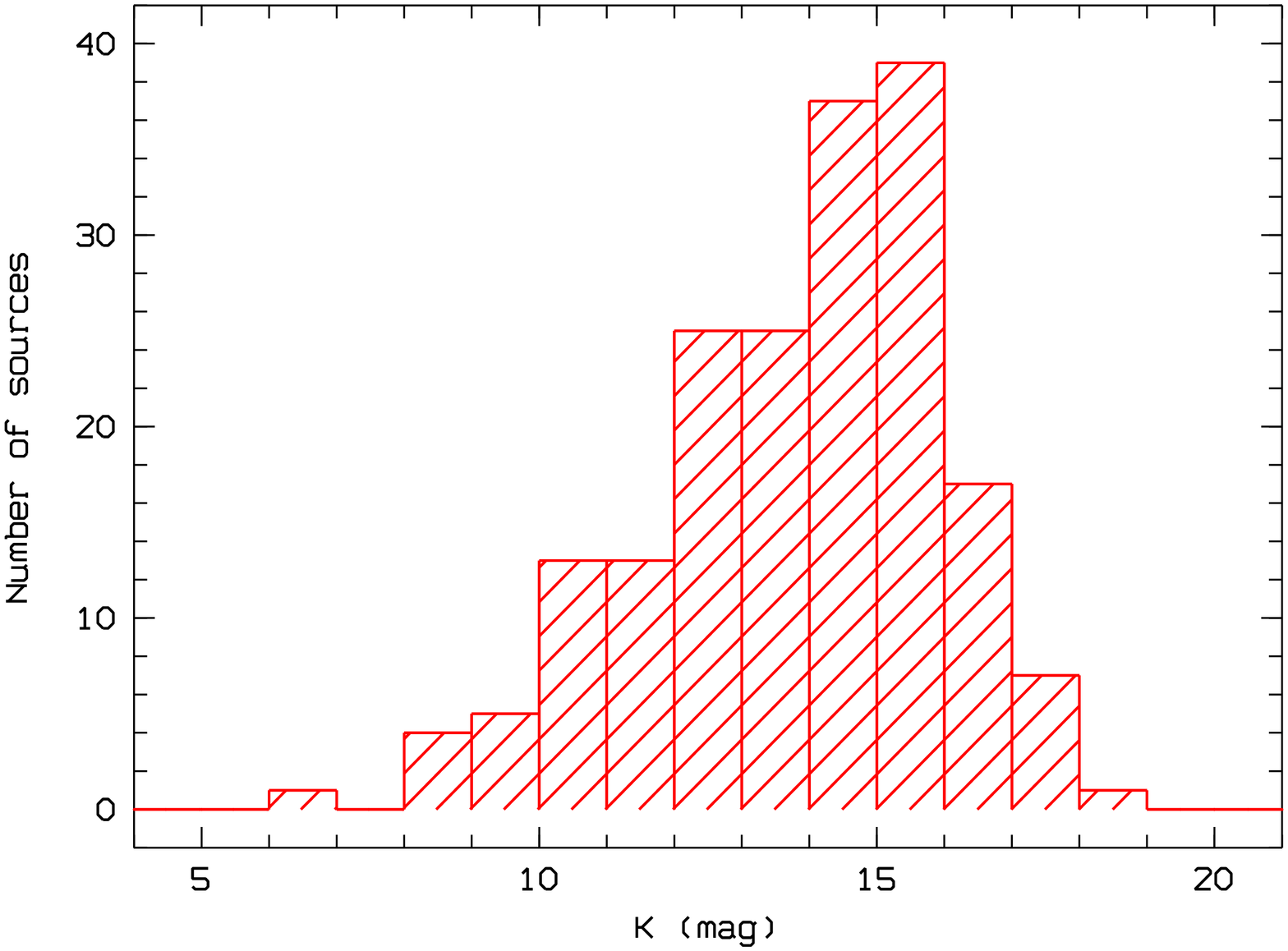}
}}
\caption{$K$ histogram of the near-IR Serpens objects.}
\label{K-histo}
\end{center}
\end{figure}

Because of  severe extinction, sensitive IR  observations are required
to sample the  stellar population embedded in the  Serpens core. Eiroa
\&  Casali (1992),  Horrobin  et  al. (1997),  Sogawa  et al.  (1997),
Giovannetti et al.  (1998), Kaas (1999), Hodapp et  al. (1996), Hodapp
(1999),  have  used  near-IR   cameras  to  study  the  young  stellar
population embedded  in the  core, within a  region $\sim  10 \arcmin$
approximately centred  in SRN. The achieved sensitivity  is $K$ $\sim
15  -  16.5$  mag.  The  images  reveal  a  relatively  large  diffuse
reflection nebula,  about $2\arcmin \times  3\arcmin$ in size,  in the
center of the  core, extending from the optical  SRN towards the South
up to the position  of the SVS 4 complex (Strom et  al. 1976, Eiroa \&
Casali 1989), some  nebulous objects and knots, and  a large number of
red  stellar objects,  which are  not visible  in  optical wavelengths
(even  up  to  0.9  $\mu$m,  e.g.  G\'omez  de  Castro  et  al.  1987,
Giovannetti  et al.  1998). A  variety of  criteria has  been  used to
identify near-IR embedded objects, e.g. near-IR excesses, variability,
association  with  nebulosity.   Fig.  \ref{K-image}  reproduces  the
$K$-band image from Kaas (1999) where many YSOs are indicated. Table 1
gives names, J2000 equatorial coordinates, and $JHK$ magnitudes of the
(up  to  now) identified  near-IR  ($JHK$)  YSOs.  The table  includes
objects  from  the  aforementioned   works,  as  well  as  from  other
references.  Further, by compiling  the table,  we have  observed that
some objects show a  non-negligible brightness difference ($\Delta K >
0.5$) between  different studies; these objects most  likely belong to
the young cluster and are therefore  included in the table. A few of the
objects listed  in the table  are located outside the  Serpens central
core; we  have chosen  to include them  for completeness. Klotz  et al.
(2004) have recently conducted a deeper ($K$ completeness limit $\sim$
19 mag.) near-IR  survey in a region $5 \arcmin  \times 10 \arcmin$ in
size. This work differs from the others because the observed area lies
towards  the  SW  from SRN,  in  a  region  where the  obscuration  is
obviously smaller than  in the core center (see Fig. 1  and 6 of Klotz
et al.  2004). Klotz et al.  suggest 14 further  candidates as Serpens
members, from which three may  be brown dwarfs. These objects are also
included  in  Table 1.  Using  Spitzer  and  Chandra data  Winston  et
al. (2007) have identified more than 100 YSOs in Serpens; some of them
have 2MASS counterparts.  These 2MASS  objects are included in Table 1
although  some   of  them  are   also  located  outside   the  central
gaseous/dusty  core.  Thus, the  total  number  of identified  near-IR
Serpens  object candidates  in  the central  core  or close  to it  is
252. The  near-IR sources  are  distributed across  the whole  Serpens
core, although  there is a  clear concentration towards the  SE clump,
either embedded in  the near-IR nebula or relatively  close to it (see
Sect. 5).

Many of the  Serpens objects appear to be binaries  or forming part of
small sub-clusters. The  SVS 4 complex is the  outstanding case, where
$\sim$ 10 objects appear to form a compact group, embedded in a common
faint nebulosity spread over a field $\sim 35\arcsec \times 35\arcsec$
in size.  SVS 20 is an example of a binary with interesting properties
(see  Sect.  8 for  a  more  detailed description  of  SVS  4 and  SVS
20).  Haisch et al.  (2002, 2004)  and Duch\^ene  et al.   (2007) have
found a  multiplicity fraction of $\sim$  35\% in a  limited sample of
Serpens Class I/flat-spectrum sources.

\begin{figure}[!ht]
\centering
\includegraphics[width=0.49\textwidth]{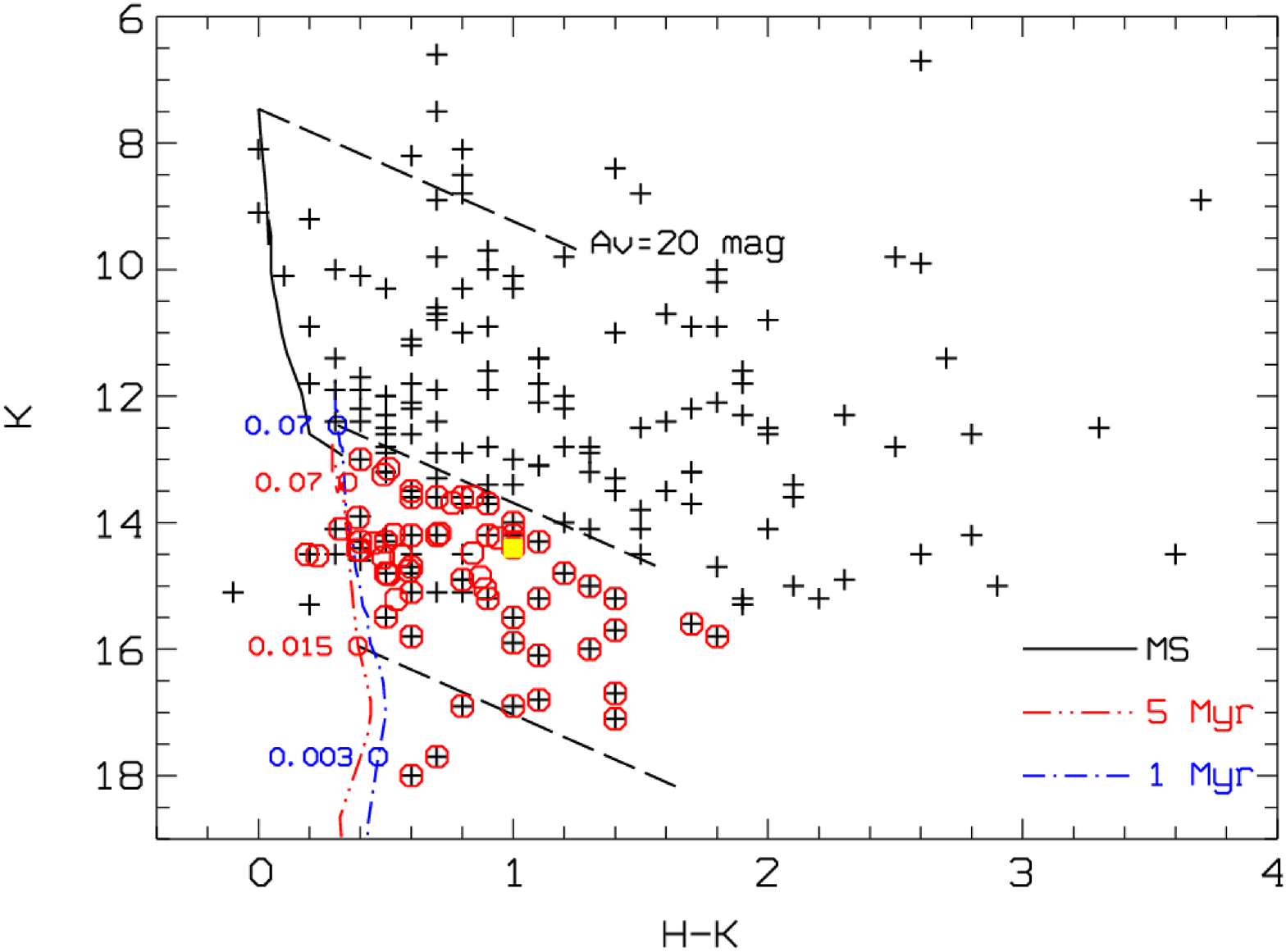}
\includegraphics[width=0.49\textwidth]{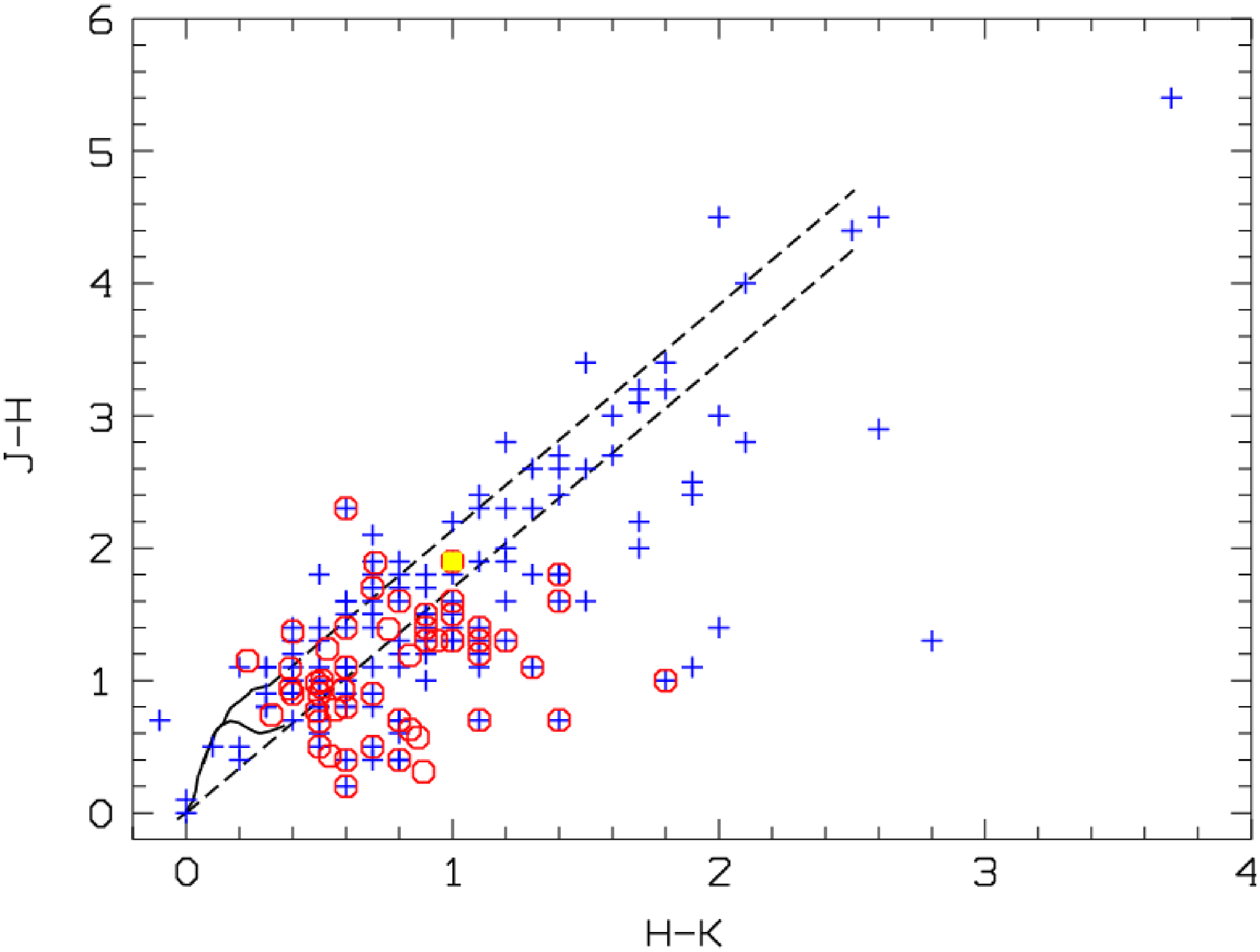}
\caption{Left:  Colour-magnitude diagrams of  the Serpens  objects. The
continuous  line represents  main-sequence stars,  while dashed-dotted
lines are  1 Myr (blue  colour) and 5  Myr (red colour)  isochrones from
Baraffe  et al.   (2003).  Dashed lines  represent reddening  vectors,
A$_V$ = 20  mag. $K$-magnitudes have been scaled to  a distance of 230
pc. Right: Colour-colour diagram  of the Serpens objects. The continuous
line represents  main-sequence and giant stars while  dashed lines are
reddening  vectors. Squares  (red colour)  in both  diagrams  are brown
dwarf  candidates (see  text); the  confirmed Serpens  brown  dwarf is
represented by the filled square (yellow colour).}
\label{magcol}
\end{figure}

Fig.   \ref{K-histo}  provides  the  histogram  of  the  apparent  $K$
magnitude of the  Serpens objects. The number of  sources increases up
to  the  sensitivity limit  of  the  near-IR  surveys, supporting  the
suggestion  of  Kaas  (1999),  and  contradicting the  presence  of  a
turnover in  the luminosity, as  suggested earlier by Eiroa  \& Casali
(1992) and  Giovannetti  et  al.    (1998).   The  histogram  in  Fig.
\ref{K-histo} does not reflect  the true luminosity and mass functions
of the Serpens near-IR cluster  since magnitudes are not corrected for
extinction, but it  does show that a non-negligible  number of Serpens
objects are  intrinsically very faint, and  consequently very low-mass
YSOs  (see  below  and  Sect.  6). Fig.  \ref{magcol}  reproduces  the
colour-magnitude ($H-K, K$) and colour-colour ($H-K, J-H$) diagrams of
the  sources. Some objects  are located  in an  ''anomalous'' position
towards  the left of  the main-sequence  line. This  is likely  due to
source    variability   (photometry    was    not   always    obtained
simultaneously), binarity, and/or scattering which can yield a blueing
in $H-K$ colour (for the most extreme case, SVS 1, see the comments by
Kaas 1999). Many sources present  an excess or lie along the reddening
line  with  large  extinction  values.  The  colour-magnitude  diagram
suggests that  some of the sources  must be reddened,  young, very low
mass  M  stars and/or  brown  dwarfs,  since  in the  colour-magnitude
diagrams they  are located  in the area  where substellar  objects are
expected. These brown dwarf candidates  are identified in Table 1.  EC
64 (GCNM 57,  BD-Ser1) was identified as a young  L0-L3 brown dwarf by
Lodieu et  al. (2002)  and two more  brown dwarf candidates  have been
suggested  by Klotz  et al.   (2004).  Winston et  al. (2007)  suggest
objects with  $m_K > 12.5$  as brown dwarfs candidates.   In addition,
Kaas et  al.  (2004)  suggest a few  brown dwarf candidates  among the
Serpens  objects on  the  basis of  ISO  data and  the estimated  low,
bolometric nebulosity ($L_{bol} \le 0.04 L_\odot$).

\begin{figure}[!ht]
\scalebox{0.5}{\rotatebox{-90}{
\includegraphics{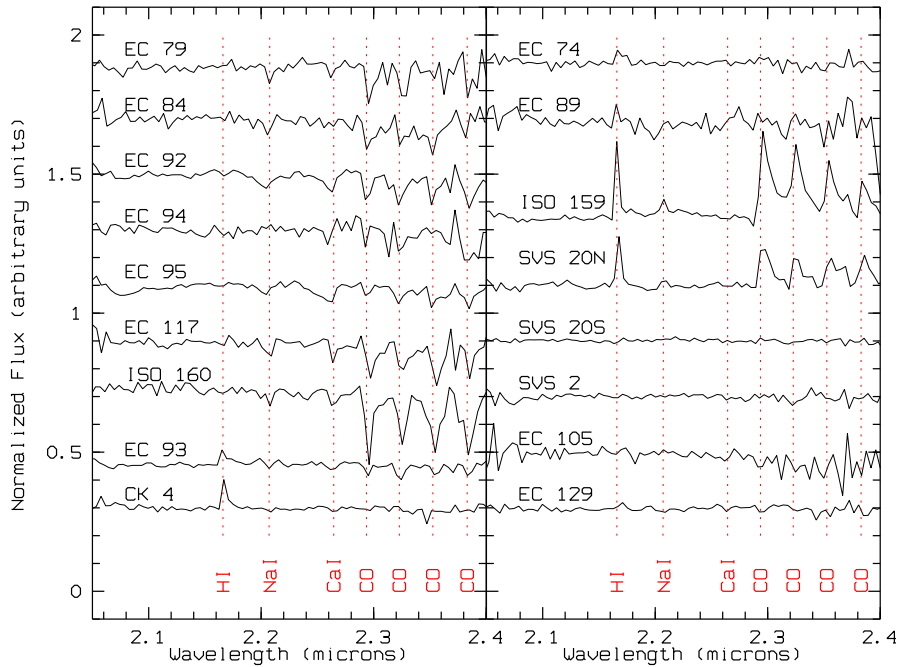}
}}
\caption{Near-IR spectra of some Serpens objects, including Class II,
flat-spectrum and Class I sources (Eiroa \& Djupvik, unpublished).}
\label{EspectrosK}
\end{figure}

Near-IR  spectra  of  some  objects   have  been  taken  by  Aspin  et
al. (1994), Casali \&  Eiroa (1996a), Preibisch (1999), Hodapp (1999),
Doppmann et al.  (2005), Covey  et al. (2006), Aspin \& Greene (2007),
Eiroa et al. (2008).  Fig.  \ref{EspectrosK} shows some representative
$K$-spectra.  The  spectra   usually  show  atomic  absorption  lines,
e.g. NaI,  CaI, as  well as CO  absorption bands, which  correspond to
late K and M spectral types,  similar to Classical T Tauri stars.  (As
noted  above, there is  a brown  dwarf spectroscopically  confirmed by
Lodieu et al.   2002).  In some cases, the  atomic and molecular bands
appear  in  emission,  most  likely  due to  circumstellar  disks;  in
addition,  the spectra of  some objects  appear to  be veiled  by dust
emission and are flat.  The spectroscopic and photometric data suggest
that  the near-IR is  mainly tracing  the Class  II population  of the
Serpens cluster. We note, however, that some flat-spectrum and Class I
sources  are detected  in the  near-IR (Kaas  et al.  2004,  Harvey et
al.  2007a,  Winston  et  al.   2007, see  next  paragraph)  and  show
photospheric lines (e.g.  Doppmann et al. 2005); even further, near-IR
counterparts  of very  young Class  0 sources  have been  suggested by
Hodapp (1999, see also Eiroa et al., 2005).

\begin{figure}[!ht]
\begin{center}
\scalebox{0.4}{
\includegraphics{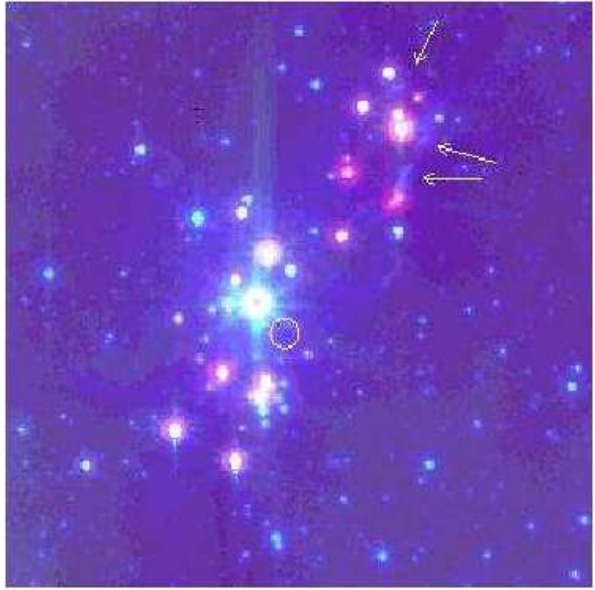}
}
\caption{Three-colour Spitzer image of the Serpens core. Colour coding is:
blue/4.5 $\mu$m, green/8.0 $\mu$m, and red/24 $\mu$m. Image taken from
Harvey et al. (2007a).}
\label{Spitzer_core}
\end{center}
\end{figure}

Serpens was observed  within the ISO central programme  in two filters
centred at  6.7 and  14 $\mu$m. 0.13  square degrees were  covered by
three overlapping maps, one centred in the core, a second one towards
the West,  and a third  one towards the  South (Kaas et al.   2004). A
total  of 421 sources  were detected,  out of  which 392  sources were
found   at  6.7  $\mu$m,   140  at   14  $\mu$m,   and  124   at  both
wavelengths. ISO  is particularly useful to detect  young sources with
no near-IR excess but with a  distinct mid-IR excess.  On the basis of
the spectral index $\alpha^{7-14}_{IR}$  the ISO sources separate into
two  distinct groups:  ``red'' sources  with IR  excess  (53 objects),
which are YSOs surrounded  by circumstellar dust, and ``blue'' sources
(71  objects), which  likely are  field stars,  although some  of them
could be Serpens  Class III objects.  Using ISO  data and near-IR data
Kaas  et  al.   (2004)  identify  20  Class  I  and  13  flat-spectrum
protostars, and 43  Class II stars (i.e. embedded  CTTs) among all the
ISO sources. Serpens is one  of the star formation regions observed by
the c2d Spitzer project (Evans et al. 2003). The Spitzer/IRAC and MIPS
observations extend  to a  large region (Harvey  et al.   2006, 2007a,
2007b, see  below). The young  stellar population of the  core, called
Cluster A by Harvey and  collaborators, is included in the analysis of
the whole Serpens cloud by Harvey  et al., while Winston et al. (2007)
specifically analyze the Spitzer core  data, extending their work to a
region   a   bit   larger   than   the  dense   central   core.   Fig.
\ref{Spitzer_core} shows a three colour, 4.5, 8.0 and 24 $\mu$m Spitzer
image  of the core  (Harvey et  al.  2007a).   By using  Spitzer data,
2MASS, and  Chandra observations, Winston  et al. (2007)  identify 229
Serpens candidate objects,  out of which 138 are  considered bona fide
YSOs: 22 Class  0/I protostars, 16 flat-spectrum sources,  62 Class II
stars, 17  transition disk candidates,  and 21 Class III  sources. The
relative number  of ISO and Spitzer  Class 0/I to Class  II sources in
Serpens is  considerably larger than  in similar young  clusters, like
$\rho$ Oph, Chamaeleon  or Taurus; in fact, the  number fraction Class
I/Class II in  the core is almost 10 times  higher than 'normal' (Kaas
et  al.  2004, Schmeja  et al.  2005, see  Winston et  al. 2007  for a
comparison  with other  regions observed  in  the context  of the  c2d
project), which  indicates that  the embedded population  is extremely
young and active.

Spitzer IRS spectra  of five Serpens objects (EC 74, EC  82, EC 90, EC
92, and CK 4) have been studied by Lahuis et al. (2007). They detected
H$_2$  S(2) and  [NeII]  emission towards  EC  74, EC  82  and EC  92,
indicative of a significant hot gas  component in the disks (T $>$ 500
K) and a [NeII] excitation mechanism through X-Rays or EUV.

\begin{figure}[!ht]
\centering
\includegraphics[width=\textwidth,draft=False]{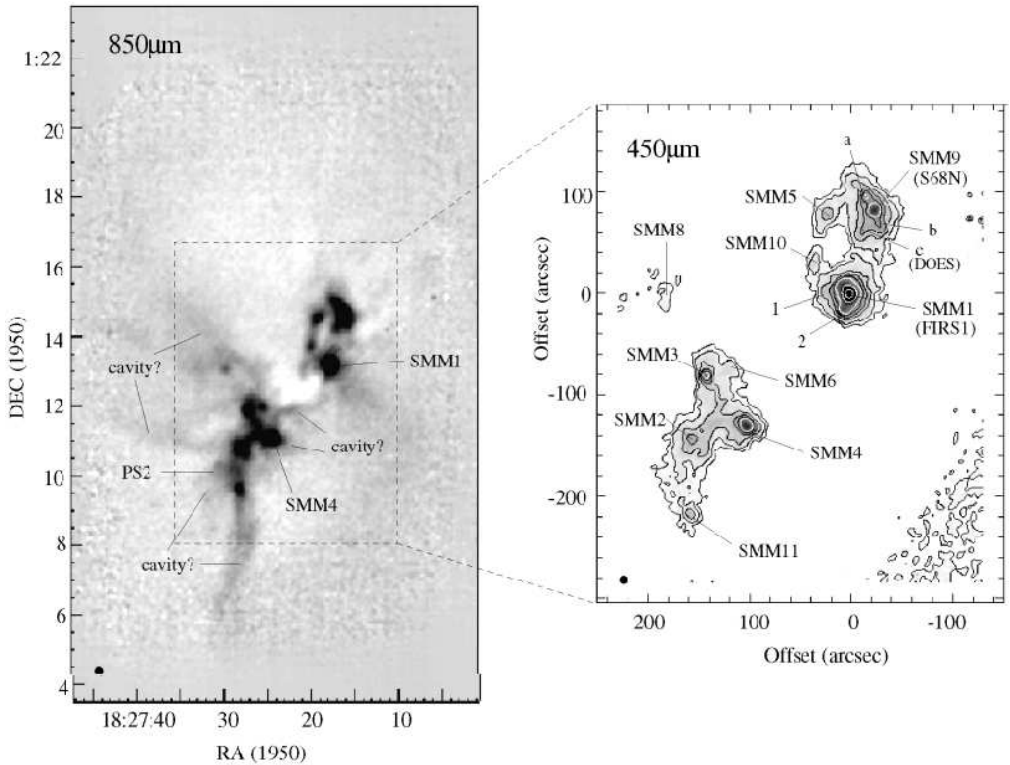}
\caption{Submillimeter images of the Serpens core. Figure taken from Davis et
al. (1999).}
\label{Davis_smm}
\end{figure}

\subsection{The Submillimeter and Millimeter Cluster: Tracing the Class 0 and
Pre-stellar Population}
\label{Class0}

About a dozen of discrete sources embedded in the cloud core have been
found by  submm observations (Casali et  al. 1993, White  et al. 1995,
Reipurth et  al.  1996, Davis et al.   1999, Enoch et al.  2007, Wu et
al. 2007).  Fig.  \ref{Davis_smm} shows the submm maps  at 850 and 450
$\mu$m of the core, after Davis  et al.  (1999). Several sources - SMM
1, SMM 5, SMM 6, SMM 9, SMM 10 - have been identified with near-IR and
ISO/Spitzer  sources  (Casali  et  al.   1993, Hodapp  1999,  Kaas  et
al. 2004, Winston et al. 2007) and some  of them - SMM 2, SMM 3, SMM 4
- with IRAS and  ISO far-IR sources (Hurt \&  Barsony 1996, Larsson et
al. 2000). An analysis of their SEDs shows that they form a cluster of
very  young,  low  to  intermediate  luminosity Class  0  objects  (or
borderline  Class 0/Class I  objects), with  dust temperatures  in the
range $\sim  20 - 50  ~K$ (Casali et  al. 1993, McMullin et  al. 1994,
2000,  Hurt  \&  Barsony  1996,  Wolf-Chase et  al.   1998,  Davis  et
al. 1999). Some sources have been modelled in terms of a stellar core,
an accretion  disk and an  extended, massive envelope  (Hogerheijde et
al. 1999, Brown  et al. 2000, Larsson et  al.  2000). Kinematic infall
signatures have been suggested for a number of the submm sources - SMM
2, SMM  3, SMM  4, SMM  9 (Hurt et  al. 1996,  Gregersen et  al. 1997,
Mardones et al. 1997, Williams  \& Myers 2000, Narayanan et al. 2002),
although at least in the case  of SMM 9 (S 68N) this interpretation is
controversial (Wolf-Chase et al. 1998).

\begin{figure}[!ht]
\scalebox{0.75}{
\includegraphics{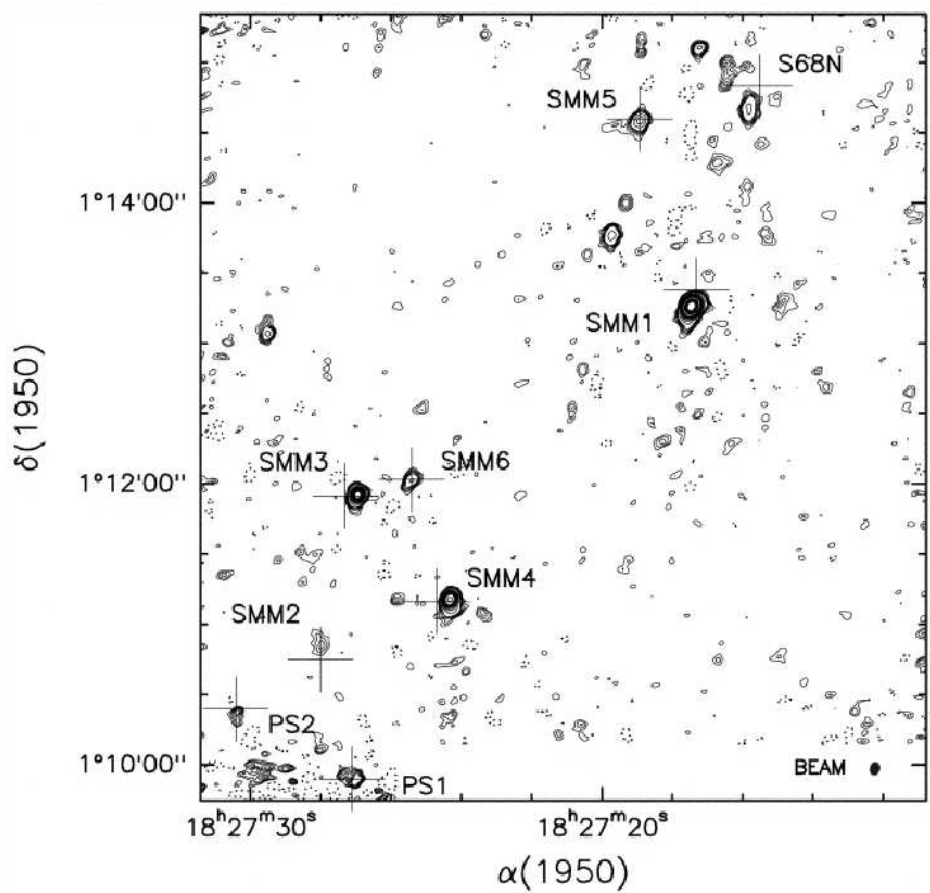}
}
\caption{3 mm continuum image from  the Serpens core. The positions of
some   submm  sources   and   far-IR  sources   (PS1   and  PS2)   are
indicated. Figure taken from Testi \& Sargent (1998).}
\label{mm}
\end{figure}

Interferometric observations at  3 mm carried out by  Testi \& Sargent
(1998) reveal  32  discrete  sources  widespread  in  the  cloud  core
(Fig. \ref{mm}), and 26 out  of 32 are probably protostellar continuum
condensations,  with masses  in the  range  0.3 -  5.0 M$_\odot$.  The
sources  are  distributed  into  two  groups:  the  more  numerous  is
associated  with the  NW core  clump and  the second  one with  the SE
clump. Seven of the 3 mm sources  in the NW clump are also detected by
Williams \&  Myers (2000). These authors  also find 8  gaseous (CS and
N$_2$H$^+$) condensations in the NW clump, with high peak temperatures
and low  velocity dispersion, 6 of which  without associated continuum
sources.   These quiescent  cores have  typical sizes  of 5000  AU and
display spectra indicative of infalling motions; they are suggested to
be   pre-protostellar   collapsing   cores.   A   further   turbulent,
contracting, starless core was also found towards the West of S 68N by
Williams \& Myers (1999). Recently, SMM 1 has been observed at 1.3 and
3.3 mm with the PdB interferometer (Fuente et al. 2007).

\subsection{Radio Continuum Observations}

Very few  observations of  the core have  been made in  the centimeter
radio  continuum (Rodr\'\i  guez  et al.   1980,  1989, Ungerechts  \&
G\"usten 1984, Snell  \& Bally 1986).  FIRS 1 is  the only source with
detailed   studies   due  to   its   unique   radio  properties   (see
Sect.  8). Smith  et al.   (1999) report  several VLA  sources  in the
region around the  SVS 4 complex.  Recently, Eiroa  et al. (2005) have
analysed 3.5 cm VLA data of the whole Serpens core. They find 22 radio
continuum sources, 16  of which seem to be  associated with YSOs.  The
radio  sources are distributed  into two  groups: one  associated with
Class 0/Class I sources in the NW  clump of the core, and a second one
towards the SE clump, where the radio continuum sources are associated
with all  kind of objects -  from Class 0  to Class II sources.  It is
likely that the radio emission from the Serpens objects is produced by
thermal jets, but  the radio emission of some sources  in the SE clump
likely originates in coronally active  PMS stars as suggested by Smith
et al.   (1999, see also Preibisch  2003, Eiroa et  al. 2005). Further
observations to estimate radio spectral indexes of the sources as well
as their  sub-arcsec structure would  be very useful to  determine the
nature of the Serpens  radio continuum objects.  Recently, Forbrich et
al. (2007) have  searched for coronal radio emission  from EC 95 using
VLBI with a non-detection result.

\subsection{X-ray Observations}

X-ray emission has been studied  by Preibisch (1998, 1999, 2003, 2004)
using  ROSAT  and XMM.   ROSAT  detected  very  strong X-Ray  emission
associated with  the flat-spectrum source EC  95, one of  the stars in
the SVS  4 group (Sect. 8).  XMM observations have  detected $\sim$ 50
sources in  a field $\approx  30 \arcmin$, several of  them associated
with  YSOs. Among the  X-ray sources,  4 are  associated with  Class I
protostars, two with flat-spectrum sources,  and at least 9 with Class
II objects.  X-ray  Class I protostars are among  the brighter Serpens
Class I objects in the $K$-band, which could mean that these YSOs have
higher bolometric luminosities and/or  suffer less extinction than the
X-ray undetected Class I objects.  The X-ray Class I objects present a
strong  flare-like  variability, with  flare  frequency  a factor  two
higher than the flare frequency in typical T Tauri stars, although the
Class II object EC74 shows a large variability in its X-ray luminosity
(a  factor  of  10). The  X-ray  emission  of  Class I  protostars  is
suggested to  be due to  the magnetic coupling between  the protostars
and their circumstellar disks.  The X-ray emission of Class II objects
is  normally ascribed  to  magnetic dynamo  processes  in the  stellar
coronae of  the PMS stars.  We point out  that 4 of the  X-ray sources
have VLA  counterparts: the  Class I object  EC 53,  the flat-spectrum
source SVS 20, and the Class II sources EC 95 and EC 117.

\begin{figure}[!b]
\scalebox{0.4}{
\includegraphics{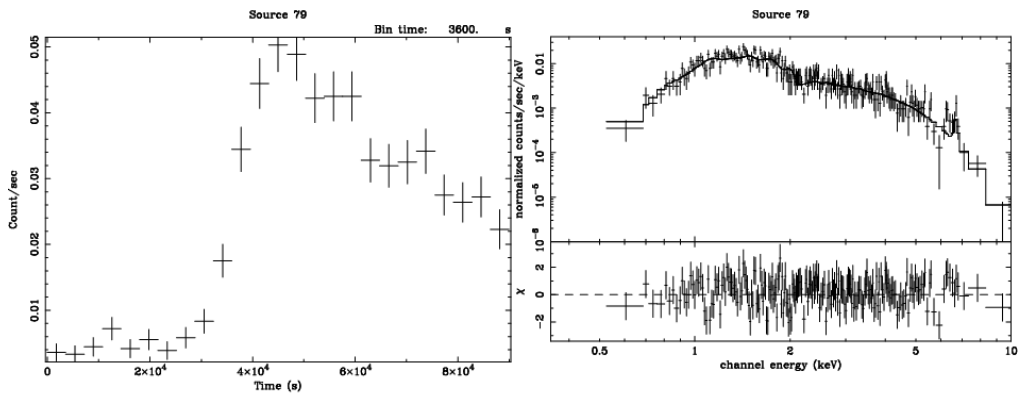}
}
\caption{Light curve  and spectra  with spectral fit  of the  Class II
object Chandra~79 (Giardino et al. 2007).}
\label{Chandra79}
\end{figure}

A  deep Chandra  X-ray  survey has  been  carried out  by Giardino  et
al. (2007), with the aim  to determine whether Class 0 protostars emit
X-rays. A total of 85 X-ray sources are detected, none associated with
Class 0 objects.  The light curves and spectra of 35 YSOs are derived.
These authors  find a trend  of decreasing absorbing  column densities
from Class  I to  Class III sources,  and some evidence  of decreasing
plasma  temperatures.  The  Chandra source  79  (ISO 393,  a Class  II
source) underwent a large,  long duration flare (Fig. \ref{Chandra79})
for which Giardino et al. (2007)  derive a semi-loop length $L = 10-12
R_\odot$, and suggest that the long flaring loop could be explained if
the flare is due to a magnetic reconnection
event of  a flux  tube linking the  star's photosphere with  the inner
ring of a  circumstellar disk. The spectrum of this  YSO shows 6.4 keV
Fe fluorescent emission compatible with reflection off a circumstellar
disk irradiated  by the  hard X-ray continuum  emission of  the flare.
Using the Chandra  data, Winston et al. (2007) find  that 60 YSOs have
detectable X-ray emission: 9 of  the X-ray sources are associated with
Class  I objects,  8  with  flat-spectrum sources,  21  have Class  II
counterparts,  2 are transition  disks. Twenty  Class III  sources are
identified solely by their X-ray emission. Comparing with the relative
numbers of Serpens  YSOs in each evolutionary class,  there appears to
be no evidence of a trend in the detection rate of sources by class.

\subsection{Outflows and Masers}
\label{Outflows}

Low spatial resolution observations by Bally \& Lada (1983) detected a
very large, $\sim 12 \arcmin  \times 17 \arcmin$, CO outflow with both
blueshifted and  redshifted components approximately  centred in SRN,
which  could  not  be  ascribed  to a  single  driven  source.  Higher
resolution  CO observations  show  widespread high  velocity gas  that
extends over the core and resolved, molecular outflows associated with
Class 0 and Class I YSOs (White  et al.  1995, also Eiroa et al. 1992,
Davis et al. 1999). NH$_3$, SiO, CS, HCO$^+$, HCN compact outflows are
associated with FIRS 1 (SMM 1), S 68N (SMM 9), SMM 3, SMM 4 and SMM 10
(Torrelles et  al. 1992,  McMullin et al.  1994, Curiel et  al.  1996,
Wolf-Chase et  al. 1998,  Hogerheijde et al.  1999, Williams  \& Myers
2000). It is interesting to note  that some of these sources also show
infall signatures (Sect. 4.2).

\begin{figure}[!ht]
\centering

\includegraphics[draft=False,width=0.85\textwidth]{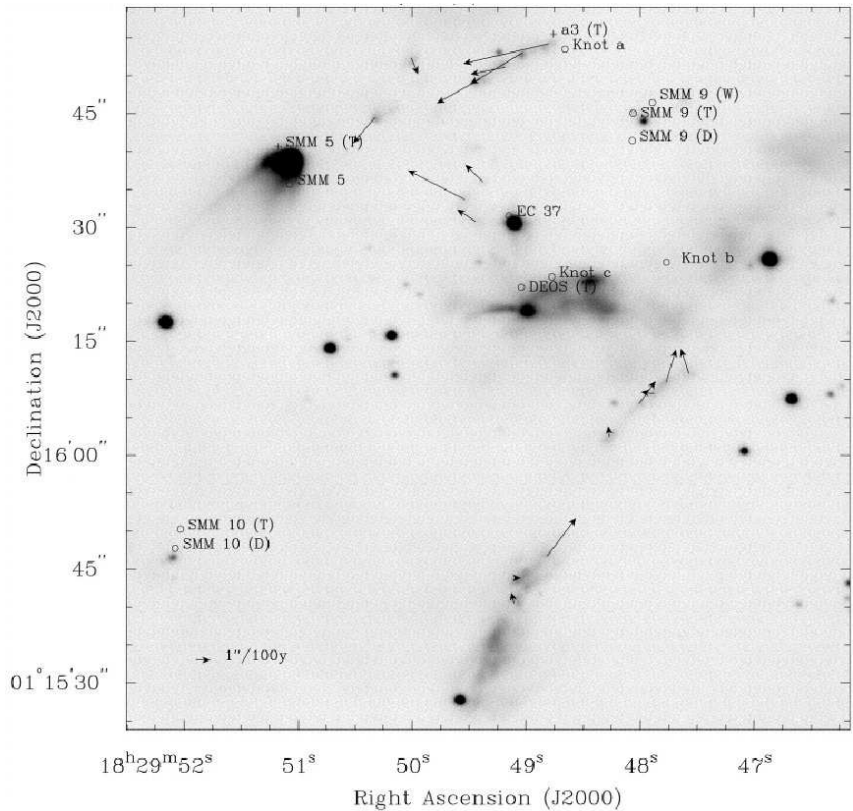}

\caption{Proper motion vectors of H$_2$ emission knots in the NW clump
of the Serpens core. Figure after Hodapp (1999).}
\label{Hodapp_fig2}
\end{figure}

An H$_2$ jet with a point-like object (EC 41) at its apex close to the
position of  FIRS 1/SMM 1 was  detected by Eiroa \&  Casali (1989, see
also Rodr\'\i guez et al. 1989), although the relationship between the
H$_2$  emission and  the Class  0  object was  unclear. Later  studies
(Curiel  et al.  1996,  Hodapp 1999)  show that  they more  likely are
unrelated  objects and  the H$_2$  jet is  not driven  by the  Class 0
object  which, on  the other  hand, is  powering radio  molecular line
outflows, and  a radio  continuum jet (see  Sect. 8). H$_2$  knots and
jets widely distributed  in the cloud core were  detected by Herbst et
al (1997,  these authors provide equatorial coordinates  of many H$_2$
condensations), Eiroa  et al.   (1997b) and Caratti  o Garatti  et al.
(2006). Hodapp (1999)  studied the proper motions of  the jets located
in the NW core clump; Fig. \ref{Hodapp_fig2} shows these results.  The
$H_2$   emission  has  been   associated  with   Class  0,   Class  I,
flat-spectrum and Class II objects - SMM 3, SMM 4, SMM 9, GCNM 53 (ISO
276), EC  37, SVS 20  and EC 105.  There is some ambiguity  in certain
cases, e.g. a jet could be  asssociated with either the Class 0 object
SMM 3 or  the Class II source EC 105. Spectral  imaging in H$_2$ using
ISO-CAM-CVF data of the Serpens core have been presented by Larsson et
al.  (2002).   These authors  also  present  ISO-LWS  maps of  several
forbidden lines in the far-IR.

\begin{figure}[!ht]
\begin{center}
\scalebox{0.7}{
\includegraphics*[20,20][700,290]{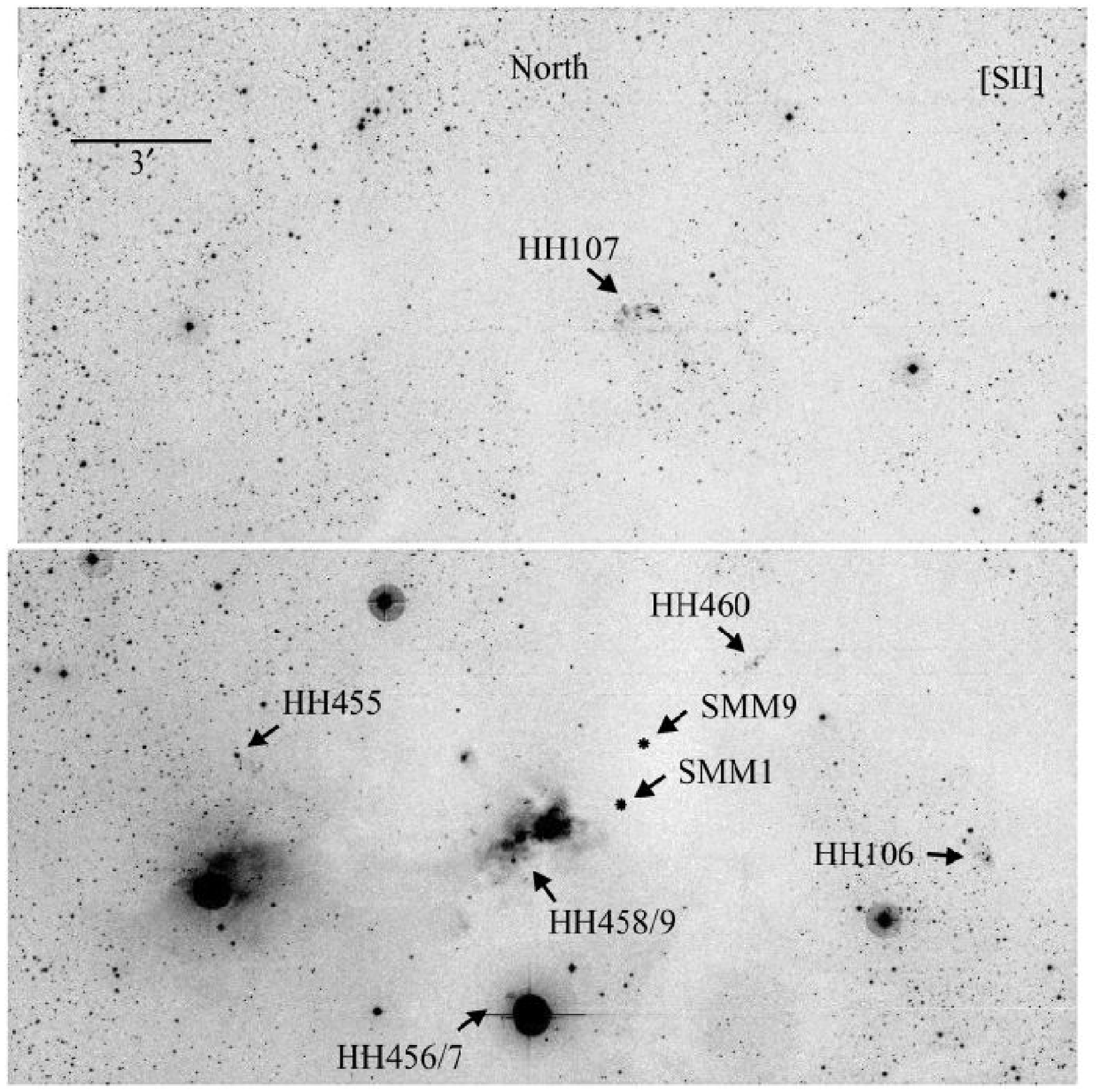}
}
\caption{Some examples of HH objects in the Serpens core. Figure after Davis
et al. (1999).}
\label{Davis_HH}
\end{center}
\end{figure}

GGD 29 was proposed as a suspected Herbig-Haro object by Gyulbudaghian
et  al.   (1978),  although  later  observations showed  it  to  be  a
reflection nebulosity with a young  star at its apex (Hartigan \& Lada
1985,  Warren-Smith   et  al.    1987).  Thus,  the   first  confirmed
Herbig-Haro objects in the Serpens  molecular cloud, HH 106/107 and HH
108/109, were discovered  by Reipurth \& Eiroa (1992).   HH 106 and HH
107 are  located towards the West  and North of the  central core (see
e.g. Fig. 2 by Ziener \& Eisl\"offel 1999). The emission line star ESO
H$\alpha$ 279a (IRAS 18296+0116), located between both HH objects, has
been suggested  as the  exciting source of  a large bipolar  flow with
both HH 106 and HH 107 at  its opposite ends (Aspin et al. 1994, Chini
et  al. 1997).  HH 108/109  are located  $\sim 2.3  \deg$ SE  from the
central  core and  likely are  excited by  the Class  I object  HH 108
IRAS/VLA 1, close to the Class 0 object HH 108 MMS (Chini et al. 1997,
2001, Siebenmorgen  \& Kr\"ugel 2000, Reipurth et  al. 2004, Connelley
et al.  2007). G\'omez de Castro (1997)  detected spectroscopically an
HH condensation 88$\arcsec$ eastwards  from SVS 2, the exciting source
of SRN.  In addition, deep CCD  H$\alpha$ and [SII] images of the core
(Davis et al.   1999) and of a 3.15 square degrees  of the large scale
cloud   (Ziener   \&  Eisl\"offel   1999)   have   revealed  many   HH
condensations,  most  of them  concentrated  in  the  core; Ziener  \&
Eisl\"offel (1999)  give equatorial coordinates  of 29 HH  objects and
condensations. Fig.  \ref{Davis_HH} shows some examples  of HH objects
in the Serpens core. Potential exciting sources of the HH objects have
been suggested, e.g. FIRS1/SMM 1, EC  105, EC 92 and EC 117.  Finally,
HH 476  is located  close to  the emission line  star group  Ser G3-G6
(Cohen \& Kuhi 1979, Ziener \& Eisl\"offel 1999).

OH maser emission associated with FIRS 1 was detected by Rodr\'\i guez
et al.   (1989, see also Clark  \& Turner 1987, Mirabel  et al. 1987).
H$_2$O  maser emission  associated  with FIRS  1  and S  68N has  been
observed  by several  groups (Blair  et  al.  1978,  Rodr\'\i guez  et
al. 1978,  1980, Dinger \&  Dickinson 1980, Palla \&  Giovanardi 1989,
Curiel et  al. 1993,  Wolf-Chase et al.   1998). Furuya et  al. (2003)
detected H$_2$O maser emission towards several Serpens Class 0 sources;
and recently  Moscadelli et  al.  (2006) have  carried out  a detailed
study of  the spatial  distribution and proper  motions of the  FIRS 1
H$_2$O maser, which is formed by two small clusters of features.

\section{Spatial Distribution of YSOs in the Core}
\label{clustering}

\begin{figure}[!ht]
\scalebox{0.625}{\rotatebox{-90}{
\includegraphics{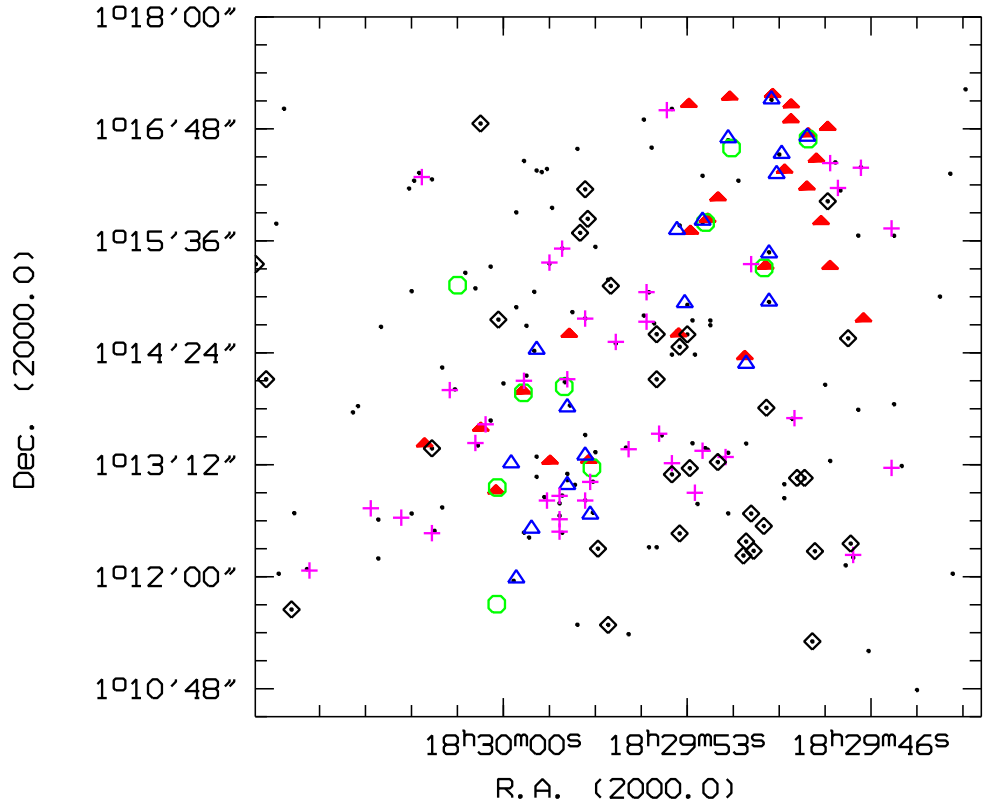}
}}
\caption{Spatial distribution  of YSOs in the Serpens  cloud core. The
plotted area  approximately corresponds to  that observed by  Testi et
al. (1998, see our Fig  \ref{mm}). Symbols mean the following: dots
(black): near-IR  sources; crosses (magenta): Class  II ISO sources;
open triangles (blue): Class  I ISO sources;  open hexagons (green):
Class 0  sources; filled  triangles (red): mm cores; diamond (black):
brown dwarf candidates.}
\label{Spatial}
\end{figure}

Eiroa \&  Casali (1992) noticed  that the Serpens near-IR  sources are
concentrated towards the core. A more detailed analysis of the spatial
distribution of  Serpens YSOs  has been made  by Testi et  al. (2000),
Kaas et  al (2004), Harvey et  al. (2007a), Enoch et  al.  (2007), and
Winston  et   al.   (2007).   The  distribution   shows  a  remarkable
subclustering, with  the youngest sources  mainly concentrated towards
the NW and SE core clumps, which  is seen by Testi et al. (2000) as an
observational   support  for   hierarchical  fragmentation   within  a
cluster-forming  core.  The  spatial distribution  of the  Serpens YSO
core cluster is shown in  detail in Fig. \ref{Spatial}, where we plot
the   distribution  of   the  sources   grouped  according   to  their
evolutionary class.  Table 2  gives a cross-correlation of the sources
detected in  the different  wavelength regimes and  their evolutionary
classes. The near-IR  cluster members are seen across  the whole core,
although their number is clearly smaller towards the field around FIRS
1  and S  68N in  the  NW clump;  brown dwarf  candidates follow  this
distribution. Most of the near-IR sources likely are Class II objects,
although  few of them  are associated  with the  youngest Class  I and
Class  0  objects (Sect. 4.1).   The  embedded Class  II
population  detected with ISO/Spitzer  is also  seen across  the whole
core, although predominantly towards the  densest part of the core and
more accentuated towards  the SE clump. Class I, Class  0 and mm cores
are exclusively seen towards the  densest parts of the two core clumps
(with only one exception). There  are approximately the same number of
Class I and Class  0 objects in each of the NW  and SE clumps, but the
number of  mm cores is clearly  larger towards the NW  clump, which is
the opposite of the spatial distribution for the near-IR and ISO Class
II  sources. These  results suggest  that star  formation  activity is
currently taking place in both core  clumps, but is more active in the
NW one, while it was more efficient or began earlier in the SE one. It
is  interesting to  recall that  both clumps  are clearly  distinct as
indicated by their  kinematics (Testi et al. 2000).  These authors and
Kaas et  al. (2004) estimate the mean  stellar/protostellar density of
the Serpens  core in the range  of $\sim$ 1000 pc$^{-3}$,  and is even
larger, several  thousands, in localized subclusters  (Eiroa \& Casali
1989, Testi et al. 2000). This high stellar density is hardly exceeded
by any other star formation  region, specially if the youngest Class 0
and  mm  cores  are  considered  (Kaas  et  al.  2004,  Testi  et  al.
2000). Winston et al. (2007)  find that the projected spacing of Class
0/I sources  is only 5000  AU, similar to  the Jeans's length  and the
radii of critical Bonnor-Ebert  spheres; they suggest that the volumes
of gas  from which these  objects accrete may overlap  and competitive
accretion  could  be  taking  place. Collisions  between  protostellar
envelopes  may be  common,  which could  explain  the widespread  high
velocity gas and overlapping outflow lobes (White et al. 1995)

\section{Luminosity and Mass Functions}
\label{massfunction}

Intrinsic luminosity  functions in young stellar  clusters are usually
based  on near-IR  magnitude  distributions; their  evaluation is  not
simple because one has  to deal with extinction estimates, variability
and   the   contribution   of   accretion  disks   to   the   observed
magnitudes. Eiroa \& Casali (1992) and Giovannetti et al. (1998) first
estimated a luminosity function on the basis of $K$-band magnitudes. A
likely  more reliable  luminosity  function of  Serpens  ISO Class  II
objects has been produced by Kaas et al. (2004). These authors compare
their  estimated Class  II  luminosity function  to pre-main  sequence
evolutionary  models  and  find   that  the  Class  II  population  is
compatible with coeval formation about  2 Myrs ago and a three-segment
power-law initial mass function, similar to other clusters like $\rho$
Ophiuchi.  Harvey et al  (2007a) find  for all  clusters in  the whole
Serpens cloud  (see next section)  that the luminosity  function peaks
around a  few times  $10^{-2}$~\Lsun and  drops to both  lower and
higher   luminosites.  Testi   \&   Sargent  (1998)   find  that   the
mass-spectrum  of the  3 mm  continuum  sources is  consistent with  a
stellar initial mass function,  supporting the idea that fragmentation
of cloud  cores determines the stellar masses.  Considering the masses
of  the different  objects  embedded  in the  Serpens  core, the  star
formation efficiency has  been estimated to be of  the order of $\sim$
3-10 \%,  although higher figures are  likely in the  densest parts of
the core clumps  (Olmi \& Testi 2000, Kaas et  al. 2004). In addition,
since the mm objects as well as the Class 0 and Class I protostars are
not older  than $\sim$  10$^5$ years, their  coexistence in  the cloud
core with the younger Class  II and near-IR cluster suggests that star
formation has proceeded in several phases (Casali et al. 1993, Kaas et
al.  2004), but see Testi et al. (2000) for an opposite view.

\section{Other Sites of Active Star Formation in the Serpens Cloud}

As already pointed out (Sect. 1) relatively few works
have  studied  the Serpens  cloud  outside  the  dense core,  although
different sites  are known  in which active  star formation  is taking
place,  e.g., Clark  (1991)  identified IRAS  sources  close to  dense
molecular material, and  the ISOCAM data of Kaas  et al. (2004) detect
some mid-IR  excess sources located several arcminutes  south from the
Serpens core.

The Herbig-Haro objects HH  108-109, located $\sim 2.3\deg$ South-East
from the  core (Reipurth \&  Eiroa 1992, Ziener \&  Eisl\"offel 1999),
form  a  bipolar  outflow  excited   by  the  Class  0/I  source  IRAS
18331--0035, which is seen at  cm wavelengths as a subarcsec radio jet
(Reipurth et  al. 2004).  The IRAS  source was detected  at 1.3  mm by
Chini et al.  (1997), as well as a second, very  cold object called HH
108 MMS; images at mm/submm wavelengths by Chini et al. (2001) show an
emission  bridge  linking both  sources.  HH  108  MMS has  only  been
detected in the  FIR and mm/submm in emission (Chini  et al. 2001) and
in absorption at 14  $\mu$m against a diffuse background (Siebenmorgen
\& Kr\"ugel 2000).  A second absorbing core, Q1, at  14 $\mu$m is also
detected.  HH  108   MMS  is  suggested  to  be   an  extremely  young
protostellar source either on the  verge of collapse or just beginning
the protostellar phase (Chini et al.  2001).

Chavarr\'\i a  et al. (1988) and de  Lara et al. (1991)  carried out a
photometric and spectroscopic study of several intermediate mass stars
associated with nebulosity, including the  Herbig Ae star VV Ser. They
also identified  13 probable H$\alpha$  emission stars. Cohen  \& Kuhi
(1979) found a close group of 4 optically visible T-Tauri stars, which
they called Ser/G3-G6, about 45$\arcmin$  to the south of the SRN (see
Fig. 1).

\begin{figure}[!ht]
\begin{center}
\scalebox{0.66}{
\includegraphics{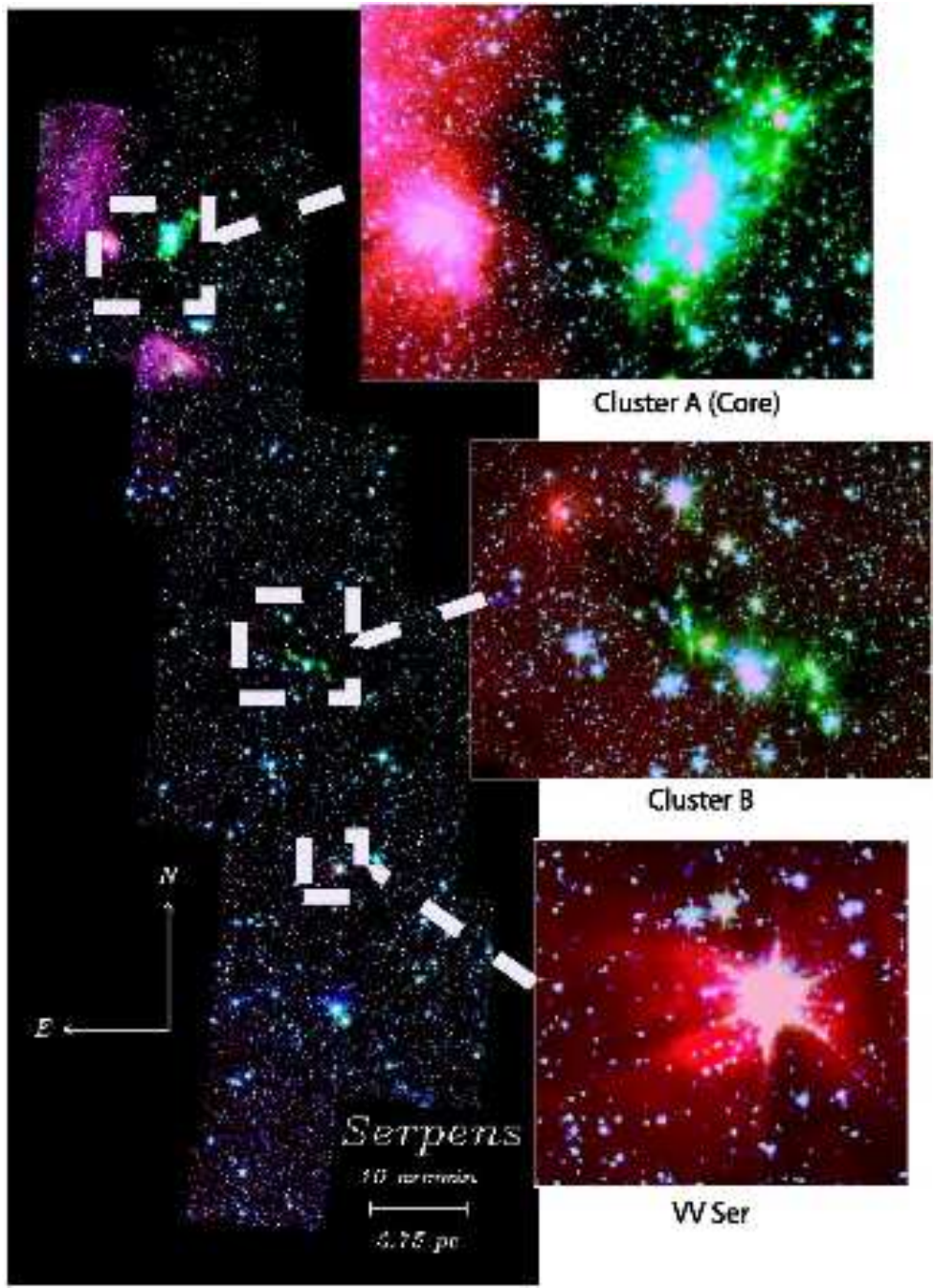}
}
\caption{Colour image  made from  Spitzer IRAC1 (blue),  IRAC2 (green),
and  IRAC4 (red)  images.  The  inset images  show  the three  Serpens
stellar clusters at higher magnification. From Harvey et al. (2006).}
\label{Harvey_Serpensall}
\end{center}
\end{figure}

The area around Ser/G3-G6 was  mapped in the ammonia 1,1 emission line
by Clark  (1990, 1991) who  found an NH$_3$  core on each side  of the
complex, Ser/G3-6ne and Ser/G3-6sw; HH  476 is close to the Ser/G3-6sw
core (Ziener \&  Eisl\"offel 1999). Those authors point  out that IRAS
18263+0027 and IRAS  18265+0028) are within about 1  and 2 arcminutes
from the Herbig-Haro object, respectively, but consider it more likely
that  its  exciting  source is  one  of  the  stars in  the  Ser/G3-G6
complex. Wu et al. (2002), on the other hand, identify IRAS 18265+0028
as the energy source of HH  476; a H$_{2}$O maser has been detected by
Persi et al.  (1994) at the position of this IRAS  source. A survey at
6.7 and 14.3 $\mu$m with ISOCAM in a 17' $\times$ 19' field centred on
Ser/G3-G6 reveals a  cluster of about 40 IR excess  sources out of 186
detections (Djupvik  et al. 2006). This population  consists mainly of
Class\,II  sources, but  also  a  handful of  Class  I sources.   Deep
follow-up imaging in  the 2.122 $\mu$m 1-0 S(1)  line of $H_2$ centred
on the  Ser/G3-6ne NH$_3$ core  shows a number  of knots and  jets not
seen in the optical. At the  position of IRAS 18265+0028 there are two
red and  bright sources, ISO-NH3-90  and 94, which probably  are Class
I/Class 0  type YSOs.  In addition, ISO-NH3-101  is a red  source with
apparent excess emission at 6.7 $\mu$m, probably indicating H$_2$ line
emission in this band.

\begin{figure}[!ht]
\centering

\includegraphics[width=\textwidth,draft=False]{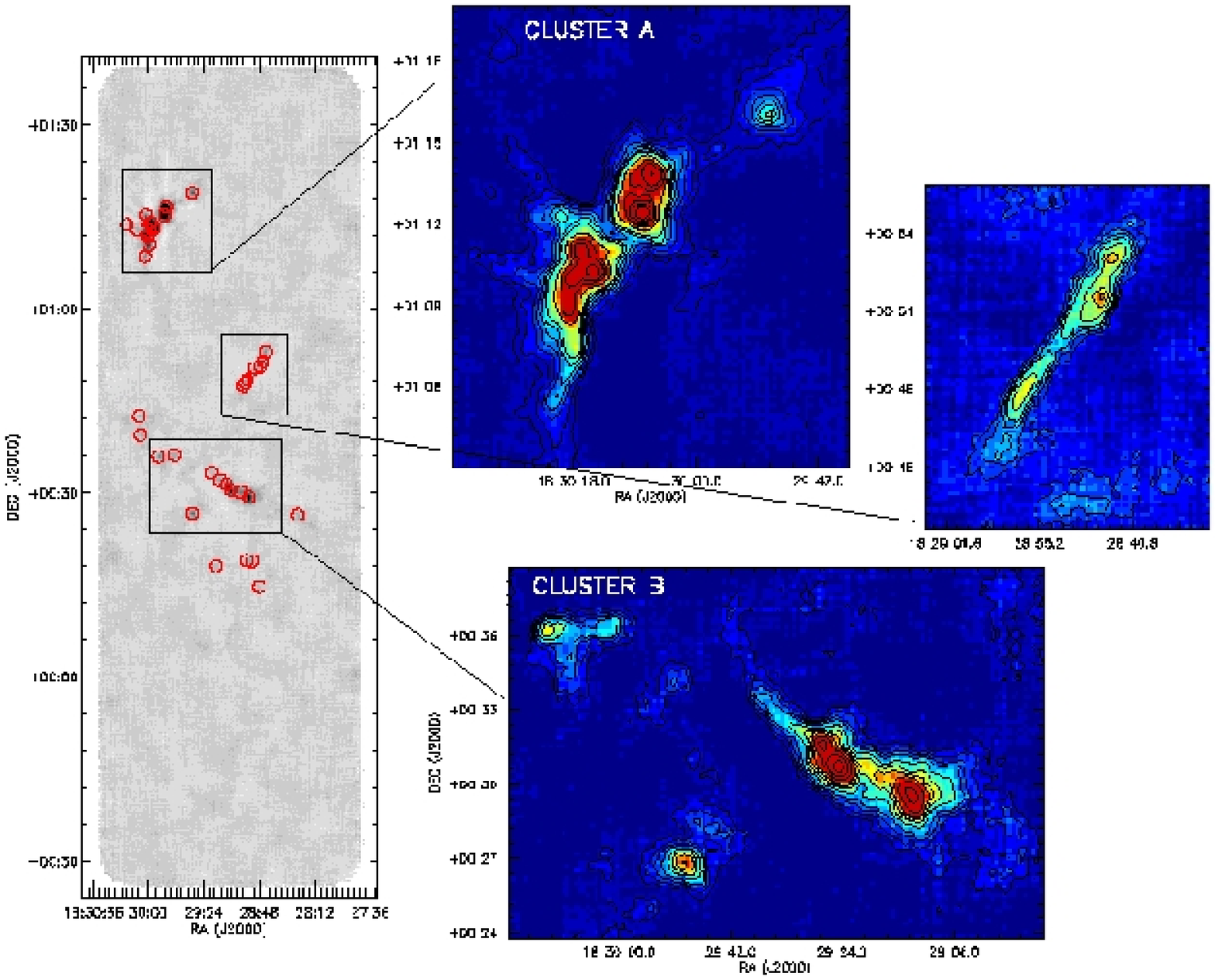}

\caption{Bolocam 1.1 mm map of Serpens with the positions of the 35 mm
sources (taken from Enoch et al. 2007).}
\label{Enoch_Fig4}
\end{figure}

Within the Legacy project  {\em From Molecular Cores to Planet-Forming
Disks} (c2d) of the Spitzer Space Telescope (Evans et al. 2003), a 0.9
deg$^2$ area in Serpens has been surveyed with IRAC and MIPS. A region
about 2 degrees in DEC and 40 arc minutes in RA has been mapped in the
4 IRAC bands (3.6, 4.5, 5.8, and 8.0 $\mu$m) and the 3 MIPS bands (24,
70 and  160 $\mu$m), the  MIPS scans being  some 20\% larger  than the
IRAC scans. This  area was chosen to follow the $A_V  = 6$ mag contour
in    the    extinction    map    of    Cambr\'esy    (1999).     Fig.
 \ref{Harvey_Serpensall}  shows  a  composite  image made  from  IRAC1,
IRAC2, and IRAC4 filter. The map includes both the Cloud Core (Cluster
A), the  Ser/G3-G6 complex (Cluster B),  and the region  around VV Ser
(Cluster C), to mention some known objects.  Spitzer/IRAC reveals more
than 250  Serpens sources distributed  into the three  mentioned areas
(Harvey et al. 2006, Harvey  et al.  2007a). This latter work presents
detailed  SEDs  of the  YSO  objects in  the  whole  area observed  by
Spitzer, traces the  YSO luminosity function down to  a limit of $\sim
10^{-3}  ~L_\odot$, and  shows  the spatial  clustering  of the  young
objects  according to  their evolutionary  classification.   They also
provide  a  table  of  Spitzer  objects  with  counterparts  at  other
wavelength ranges.  Flaherty et al.  (2007) have  studied the infrared
extinction law  in the  IRAC bands  and the 24  $\mu$m MIPS  band. The
extinction law is similar to that of other regions but it differs from
the diffuse ISM, which could  be due to different dust properties.  In
addition, Harvey et  al. (2007b) find a close  correlation between the
coolest  dust detected  at 160  $\mu$m  and the  dust associated  with
optical extinction. The region observed  by Spitzer has been mapped at
1.1 mm  by Enoch  et al. (2007).  They find  35 sources, most  of them
located in the Cluster A and Cluster B (Fig. \ref{Enoch_Fig4}). Those
authors correlate  the mm  emission with the  extinction map  and find
that  most of the  mm sources  lie within  regions of  high extinction
(A$_V \, > \, 10$ mag).

\begin{figure}[!p]
\centering
\includegraphics[width=\textwidth,draft=False]{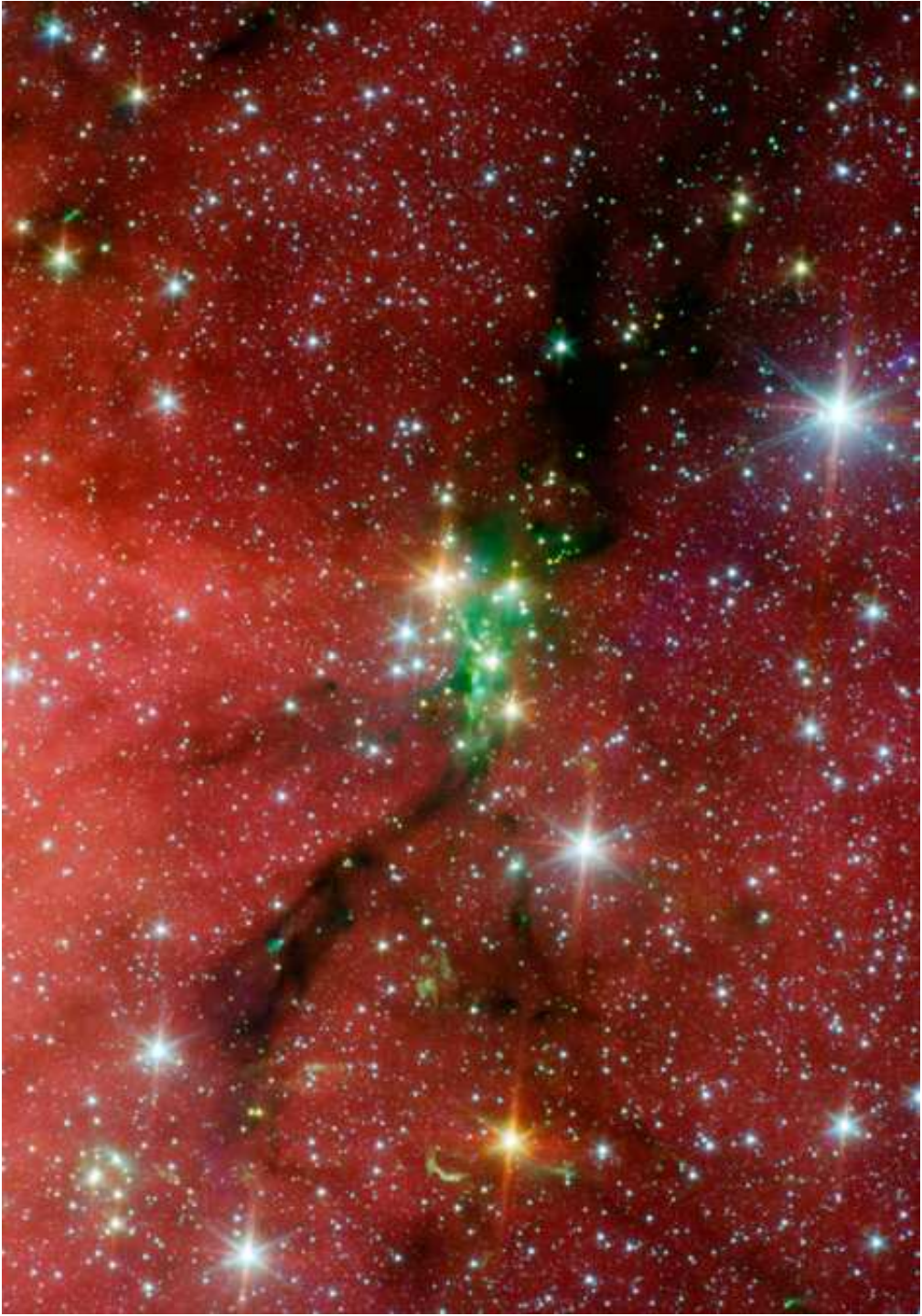}
\caption{Colour-composite of the new Serpens South cluster made from IRAC1
(blue), IRAC2 (green), and IRAC4 (red) images. From Gutermuth et al. (2008).}
\label{Serpens_South}
\end{figure}

In the  context of the Spitzer  Gould Belt Legacy Survey,  a new young
embedded   cluster  (Fig.   \ref{Serpens_South})  associated   with  a
filamentary dark  cloud, located 3$\deg$  to the South of  the Serpens
cloud core, has  been discovered by Gutermuth et  al.  (2008).  Radial
velocities  from  molecular line  observations  of  the Serpens  South
complex carried  out by the same authors  are the same as  that of the
Serpens core, suggesting a common distance. The cluster consists of at
least 54 Class I protostars and 37  Class II sources and it has a mean
surface  density  of  430  pc$^{-1}$. The  median  projected  distance
between nearest neighbor YSOs is 13$\farcs$2 or 3700 AU - Gutermuth et
al. (2008) assume a distance of 260 pc.  The large fraction of Class I
protostars in  the Serpens South  cluster, larger than in  the Serpens
cloud  core  cluster, and  the  high  density  suggest a  very  recent
initialization  of star  formation  and a  high  star formation  rate,
$\sim$ 90 $M_\odot$ Myr$^{-1}$  (assuming a typical protostellar phase
lifetime    of   2$\times$10$^5$    yr   and    0.5    $M_\odot$   per
source). Doubtless, Serpens South will concentrate large observational
efforts in  the next future,  searching for Class~0  objects, submm/mm
dusty cores, outflows, etc.

\section{Objects of Particular Interest}
\label{Interesting Objects}
\subsubsection{The Gaseous Condensation S 68 NW:} Williams \& Myers (1999)
detected a  compact, starless  core in CS  (2-1) and  N$_2$H$^+$ (1-0)
with  turbulent widths and  infall asymmetry  in the  NW clump  of the
Serpens core  close to S  68N, which was  designated S 68NW.  The line
width was confirmed  by Olmi \& Testi (2000). This  core is likely the
youngest, individual evolutionary structure detected in Serpens.

\subsubsection{SMM 4:} This source is the brightest Class 0 object in the SE
core  clump  (Casali  et  al.   1993,  Hurt  \&  Barsony  1996).  Both
Ward-Thompson \& Buckley (2001) and Narayanan et al. (2002) showed that
it possesses blue asymmetric line profiles in the optically thick main
isotopes of HCO$^+$ and CS, and  conclude that SMM 4 has an infalling,
rotating envelope.

\subsubsection{FIRS 1 (IRAS 18273+0113, SMM 1):}

FIRS 1 (Harvey et al. 1984) is the most luminous source in the Serpens
core.   It is a  Class 0  object of  $\sim 50  \, L_\odot$  (Casali et
al. 1993, Hurt \& Barsony  1996, Larsson et al. 2000). Radio continuum
observations detect  a triple radio  jet with a  non-thermal component
which  is interpreted  as a  Herbig-Haro precursor  (Rodr\'\i  guez et
al. 1989, Curiel  et al. 1993, McMullin et  al.  1994). The outflowing
clumps in  the jet have  proper motions with tangential  velocities of
order 200 km/s,  indicating a dynamical age of  only 60 years.  H$_2$O
maser  and $OH$  emission  are also  detected  close to  FIRS 1  (e.g.
Moscadelli et  al.  2006).  Molecular outflows  in different molecules
are also present (White et al.  1995, Torrelles et al. 1992, Curiel et
al. 1996,  Williams \&  Myers 2000, McMullin  et al.  2000).  An $H_2$
S(1) jet emerges  from the centre of FIRS 1, though  it is most likely
unrelated  to the  radio  sources  (Eiroa \&  Casali  1989, Curiel  et
al. 1996, Hodapp 1999).

\subsubsection{SVS 4:}

Testi et al.  (2000) point out that star formation in the Serpens core
is very  inhomogeneous, occurring in  various sub-clusters. The  SVS 4
complex (Eiroa \&  Casali 1989) is a sub-cluster  of $\sim 10$ sources
situated some 2$\arcmin$  South of the SRN, containing  a mix of Class
I, flat-spectrum and  Class II sources (Kaas et  al. 2004, Pontoppidan
et al. 2004, Haisch et al.   2006). Some of the objects are associated
with  radio  continuum sources  (Smith  et  al.   1999, Eiroa  et  al.
2005). It  is one of  the densest young stellar  (sub-)clusters known,
with a stellar mass density of $\sim 10^5 ~M_\odot ~pc^{-3}$ (Eiroa \&
Casali 1989, Pontoppidan et  al.  2004).  Preibisch (1998, 1999) finds
EC 95 (SVS  4-9) to be one of the brightest  X-ray sources known among
young stars.   Its high X-ray  luminosity, $\sim 4  \times 10^{31}$erg
s$^{-1}$, might be due to a very active magnetic corona, a fact partly
supported by the radio spectral  index (Smith et al. 1999, Forbrich et
al. 2007).  The infrared data places  it well above  the main sequence
with a  very young age  in the  range $\sim 2  - 4 \times  10^5$ years
(Preibisch  1999, Pontoppidan  et al.  2004).   A spectrum  of the  CO
bandhead at 2.29  $\mu$m in EC 95 indicates $v  \sin(i) \sim$ 20-80 km
s$^{-1}$  (Casali  \&  Eiroa  1996b), and  reasonably  rapid  rotation
consistent with  enhanced X-ray  emission; more recently,  Doppmann et
al. (2005)  have measured $v  \sin(i)$ = 56  km s$^{-1}$.  EC  80 (SVS
4-3),  and EC 89  (SVS 4-6)  are also  X-ray sources  (Preibisch 2004,
Giardino et al. 2007).

\subsubsection{SVS 20:}

SVS  20  (EC  90) is  the  brightest  K  band  source in  the  Serpens
cluster.    The    source    contains    a    number    of    peculiar
features.  Observations by  Eiroa et al. (1987)  revealed it  to  be a
1.6$\arcsec$  separation binary  and the  total luminosity  and X-rays
suggest  two  stars of  mass  1-4  M$_\odot$  (Preibisch 2003).   Both
components  have   very  similar   near  to  mid-IR   spectral  energy
distributions,  although  they have a brightness  difference of
$\Delta m  \sim $ 1.5 mag (Eiroa  et al. 1987, Eiroa  \& Leinert 1987,
Haisch et  al. 2002) and are  flat-spectrum sources (Kaas  et al. 2004,
Haisch et al. 2006). The  $K$ band spectrum of the brighter component,
SVS  20S, is featureless,  while SVS  20N has  $Br\gamma$ and  the 2.3
$\mu$m  CO  bands in  emission  (Fig.  \ref{EspectrosK}).  A  nebulous
ring-like structure  possessing spiral arm-like  features, $\sim$ 5000
$\times$ 3000 AU projected onto  the sky, contains shock excited H$_2$
emission  (Eiroa et  al.  1997a, see Fig \ref{HAWK}). The  correct  
interpretation of  the
nebular  morphology  is a  matter  of  debate.  Eiroa et  al.  (1997a)
interpret  the  emission as  a  circumbinary  disk,  while studies  in
polarized IR  radiation led Huard  et al. (1997)  to describe it  as a
cavity excavated by an outflow.

\subsubsection{Brown Dwarfs:}

The near-IR  object EC 64 (GCNM  57) has been identified  by Lodieu et
al. (2002)  as a  young L0-L3  brown dwarf (BD-Ser  1). It  shows deep
water absorption  consistent with an effective temperature  of 2800 K,
and an estimated mass of 0.05 M$_\odot$.  Klotz et al. (2004), Kaas et
al.  (2004),  Harvey et al.  (2007a)  and Winston et  al.  (2007) have
suggested further  brown dwarf candidates.  Furthermore,  on the basis
of  the near-IR  diagrams  Eiroa  et al.  (2006)  suggested $\sim$  40
near-IR Serpens objects as good  brown dwarf candidates, which are now
increased to  more than  60 candidates, considering  the Spitzer/2MASS
sources.   Many  of  them  show  near-IR  excess  denoting  remarkable
circumstellar disks. These objects are identified in Table 1.

\subsubsection{Outbursts and Disappearances. EC 81 and DEOS:}

\begin{figure}[!ht]
\begin{center}
\plotfiddle{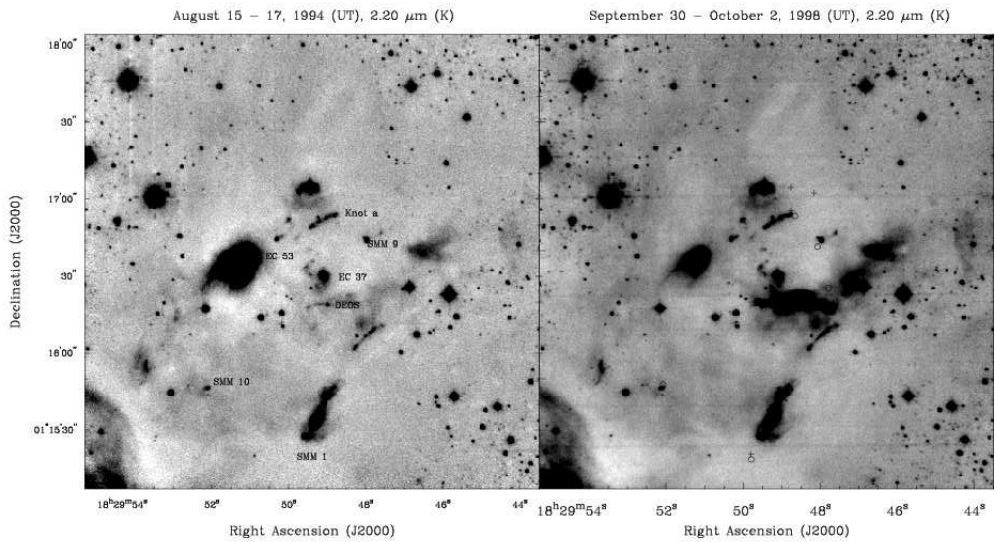}{6.9cm}{270.0}{51.0}{51.0}{-210.0}{276.0}
\caption{Combined $K$ band images of the DEOS region in 1998 (right) and
1994 (left). DEOS and its associated luminosity to the east appear much
brighter in 1998 than in 1994. Taken from Hodapp (1999).}
\label{DEOS}
\end{center}
\end{figure}

EC 81 was  a $K$ = 12.2 mag. source, with  some cometary nebulosity in
the original EC listing (Eiroa  \& Casali 1992).  Yet it was invisible
in the images reported in Horrobin et al. (1997) which show it to have
faded by more than 3.8 magnitudes  over a period of 3 years. Harvey et
al.   (2007a) have found several knots  of  emission close  to the  nominal
position of EC  81 and have suggested that this source may  have been a small
clump of excited  gas. By contrast, Hodapp et  al.  (1996) reported an
FU Orionis-type outburst in a faint source (DEOS), which brightened by
more than  4 magnitudes in less  than 9 months.  Hodapp (1988) carried
out  a first  photometric  monitoring of  DEOS  (Fig. \ref{DEOS})  and
showed  significant  changes  in  the  spectrum of  the  source.  More
recently,  K\'osp\'al et al.  (2007) has  monitored this  source, also
called OO  Serpentis, from 1996 to  2006 using ISO,  Spitzer and near-
and  mid-IR ground-based  data. Their  data  show that  the object  is
fading and is likely to reach its pre-outburst brightness around 2011;
those authors suggest DEOS is a Class I object with an age of $< 10^5$
yr surrounded by an accretion disk and a dense envelope.

\subsubsection{VV Ser:}
\begin{figure}[!t]
\begin{center}
\scalebox{0.4}{
\includegraphics{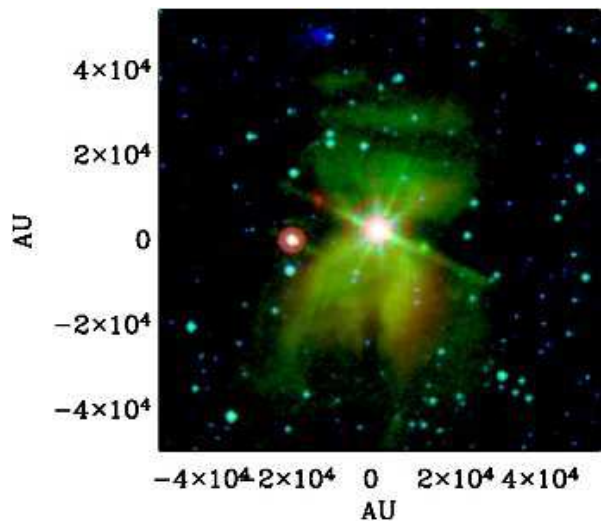}
}
\caption{Spitzer composite image of VV Ser. Colours mean: 4.5 $\mu$m (blue),
8.0 $\mu$m (green), 24 $\mu$m (red). Taken from Pontoppidan et al. (2007c).}
\label{VVSer}
\end{center}
\end{figure}
VV Serpentis  (Chavarr\'\i a et  al. 1988) has  attracted considerable
attention in  many studies of HAe  stars. In spite of  this, its basic
stellar parameters  remain rather uncertain: spectral  type between B1
and A3, luminosity class III  to V, rotational velocity between 85 and
230 km s$^{-1}$.  Ripepi et al.  (2007) provide a very good summary of
the  current  published values.   Photometrically  VV  Ser shows  deep
occultations  from circumstellar  matter (Herbst  \&  Shevchenko 1999,
Rostopchina  et   al.   2001)   together  with  $\delta$   Scuti  type
pulsations, suggestive of  a stellar mass close to  4 M$_\odot$ and an
effective temperature significantly lower  than the value based on the
empirical  spectral  types  (Ripepi  et al.   2007).   Interferometric
measurements resolve  the near-IR emission (Eisner et  al. 2003, 2004)
which is  interpreted as arising from  the inner parts of  a disk with
grains    of   different    sizes    (see   also    Isella   et    al.
2006). Double-peaked 4.7 $\mu$m CO emission is also observed (Blake \&
Boogert 2004).   Searches for PAH emission features  have been carried
out by Geers et al. (2006,  2007). Spitzer IRAC and MIPS images reveal
the  circumstellar  disk (Pontoppidan  et  al.   2007c)  and a  large,
bipolar  nebulosity centred  on VV  Ser (Fig.  \ref{VVSer}),  which is
interpreted    as    due   to    quantum-heated    PAHs   and    small
carbonaceous/silicates  of  $\sim$   500  atoms  (Pontoppidan  et  al.
2007b).

\acknowledgements  It  is  a  pleasure  to thank  R.   A.   Gutermuth,
P.M. Harvey, and S.T.  Megeath for providing the Spitzer results prior
to  publication. C.E.   is supported  in  part by  Spanish Grants  AYA
2005-00954 and CAM Ref. S-0505/ESP/0361.

\begin{table}[!ht]
\label{NIR-sources}
\caption{Near-IR sources embedded in the Serpens clouds.}
\smallskip
\begin{center}
{\small
\begin{tabular}{rllrrrcl}
\tableline
\noalign{\smallskip}
No. &$\alpha_{2000}$&$\delta_{2000}$&$J$  &$H$  &$K$   &  &Name\\
\noalign{\smallskip}
\tableline
\noalign{\smallskip}
  1 &18 29 04.3 &01 18 39 &16.7 &16.0 & 15.5 &! &WMW72  \\
  2 &18 29 05.8 &01 13 23 &16.3 &15.8 &      &  &WMW97  \\
  3 &18 29 08.2 &01 05 45 &10.7 &10.2 & 10.1 &  &WMW124 \\
  4 &18 29 13.0 &01 07 01 & 8.8 & 7.3 &  6.6 &  &WMW122 \\
  5 &18 29 14.3 &01 07 30 &14.6 &13.7 & 13.2 &! &WMW53  \\
  6 &18 29 16.4 &01 08 22 &16.3 &15.3 & 14.8 &! &WMW136 \\
  7 &18 29 20.4 &01 21 04 &12.8 &11.3 & 10.3 &  &WMW104 \\
  8 &18 29 21.1 &01 24 07 &16.3 &15.4 & 14.8 &! &WMW57  \\
  9 &18 29 22.7 &01 10 33 &12.2 &11.1 & 10.9 &  &WMW221 \\
 10 &18 29 31.6 &01 15 06 &16.0 &14.7 & 14.2 &! &WMW224 \\
 11 &18 29 31.9 &01 18 42 &11.6 & 9.8 & 8.4  &  &ESO H$\alpha$279a/WMW36 \\
 12 &18 29 31.8 &01 18 34 &17.0 &12.5 & 9.9  &  &ESO H$\alpha$279b$^\#$ \\
 13 &18 29 32.2 &01 12 40 &16.3 &15.7 & 14.9 &! & WMW75 \\
 14 &18 29 33.1 &01 17 17 &11.4 &10.3 & 10.0 &  & WMW225\\
 15 &18 29 33.4 &01 08 24 & 9.8 & 9.4 & 9.2  &  & WMW82 \\
 16 &18 29 34.1 &01 17 29 &14.8 &12.5 & 11.4 &  &KCML14 \\
 17 &18 29 35.2 &01 13 11 &15.5 &14.5 & 13.7 &! &WMW91  \\
 18 &18 29 37.5 &01 14 54 &14.5 &13.1 & 12.4 &  &KCML13 \\
 19 &18 29 37.6 &01 11 18 &11.2 & 9.3 & 8.5  &  &WMW134 \\
 20 &18 29 37.7 &01 11 30 &11.5 & 9.6 & 8.8  &  &WMW127 \\
 21 &18 29 38.2 &01 09 18 &15.9 &14.8 & 14.5 &! &WMW64  \\
 22 &18 29 39.8 &01 17 12 &15.3 &12.9 & 11.8 &  &KCML12 \\
 23 &18 29 39.9 &01 17 56 &14.5 &12.8 & 11.9 &  &WMW94  \\
 24 &18 29 41.4 &01 07 39 &11.6 &10.6 & 9.7  &  &STGM3/WMW103 \\
 25 &18 29 41.5 &01 10 05 &12.9 &11.1 & 10.3 &  &WMW176  \\
 26 &18 29 41.9 &01 17 13 &18.6 &15.4 &13.7  &  &KCML11  \\
 27 &18 29 42.2 &01 20 21 &13.3 &12.5 &12.0  &  &WMW93   \\
 28 &18 29 42.4 &01 12 02 &16.7 &14.1 &12.8  &  &KCML6/WMW215 \\
 29 &18 29 42.5 &01 16 19 &21.1 &17.1 &15.0  &  &KCML10   \\
 30 &18 29 43.1 &01 15 00 &17.0 &15.2 &14.2  &  &EC1/KCML9\\
 31 &18 29 43.8 &01 07 21 &13.7 &12.5 &11.6  &  &STGM2/WMW65\\
 32 &18 29 43.8 &01 10 47 &13.6 &11.8 &11.0  &  &KCML5      \\
 33 &18 29 44.3 &01 04 54 & 8.5 & 6.6 & 5.4  &  &WMW71    \\
 34 &18 29 44.4 &01 13 11 &13.3 &12.1 &11.7  &* &EC11/WMW100\\
 35 &18 29 44.7 &01 13 51 &17.0 &14.0 &12.4  &  &EC12/KCML8 \\
 36 &18 29 44.7 &01 15 39 &     &15.1 &13.5  &* &EC13       \\
 37 &18 29 44.9 &01 19 57 &14.7 &13.6 &12.9  &  &STGM34     \\
 38 &18 29 45.7 &01 11 12 &16.2 &14.4 &13.4  &  &KCML4      \\
 39 &18 29 46.0 &01 16 23 &     &13.7 &11.8  &  &EC21       \\
 40 &18 29 46.1 &01 13 47 &     &13.9 &12.1  &  &EC22       \\
 41 &18 29 46.1 &01 15 39 &     &     &17.3  &  &K1 	      \\
 42 &18 29 46.2 &01 08 56 &16.2 &15.3 &14.8  &! &BD-Ser2    \\
 43 &18 29 46.3 &01 10 25 &13.2 &12.3 &11.9  &  &WMW79      \\
 44 &18 29 46.3 &01 12 12 &13.3 &11.7 &11.1  &  &EC23       \\
 45 &18 29 46.4 &01 12 21 &     &14.7 &14.3  &! &EC25/K2    \\
 46 &18 29 46.6 &01 12 07 &     &     &14.1  &  &K3         \\
 47 &18 29 46.5 &01 14 33 &18.4 &17.7 &16.9  &! &K4          \\
 48 &18 29 46.8 &01 16 08 &     &     &14.3  &* &EC26        \\
 49 &18 29 47.0 &01 16 26 &     &     &13.4  &* &EC28/WMW5   \\
 50 &18 29 47.2 &01 07 16 &15.9 &14.8 &14.4  &! &WMW68       \\
 51 &18 29 47.2 &01 13 14 &     &     &16.1  &  &K5          \\
\noalign{\smallskip}
\tableline
\end{tabular}
}
\end{center}
\end{table}

\setcounter{table}{0}

\begin{table}[!ht]
\caption{Continuation.}
\smallskip
\begin{center}
{\small
\begin{tabular}{rllrrrcl}
\tableline
\noalign{\smallskip}
No. &$\alpha_{2000}$&$\delta_{2000}$&$J$  &$H$  &$K$   &  &Name\\
\noalign{\smallskip}
\tableline
\noalign{\smallskip}
 52 &18 29 47.2 &01 22 35 &11.8 &10.5 &10.10 &  &WMW199      \\
 53 &18 29 47.3 &01 16 01 &18.6 &17.6 &15.8  &! &K6/BD-Ser3   \\
 54 &18 29 47.4 &01 14 03 &     & 18.1&14.5  &  &EC29,GCNM1   \\
 55 &18 29 47.8 &01 12 16 &17.1 &15.7 &15.1  &+,! &EC30,GCNM2  \\
 56 &18 29 47.9 &01 11 18 &16.5 &15.4 &14.1  &  &STGM5         \\
 57 &18 29 48.1 &01 16 45 &     &     &16.5  &  &SMM9-IR/WMW23 \\
 58 &18 29 48.2 &01 13 04 &18.9 &18.4 &17.7  &! &K7            \\
 59 &18 29 48.5 &01 13 04 &16.2 &14.8 &14.2  &+,! &EC32,GCNM8   \\
 60 &18 29 48.7 &01 13 42 &17.6 &14.9 &13.5  &  &EC33/GCNM13/K8/\\
    &           &         &     &     &      &  &KCML7/WMW106\\
 61 &18 29 49.0 &01 12 51 &     &     &16.6  &  &K9              \\
 62 &18 29 49.0 &01 13 00 &15.5 &13.7 &12.8  &  &EC36/GCNM18     \\
 63 &18 29 49.1 &01 16 21 &     &     &10.8  &  &DEOS/K4\_5/WMW11\\
 64 &18 29 49.2 &01 16 32 &     &     &13.2  &  &EC37/GCNM19/\\
    &           &         &     &     &      &  &WMW6\\
 65 &18 29 49.5 &01 12 04 &16.2 &14.8 &14.4  &! &WMW218          \\
 66 &18 29 49.5 &01 17 07 &     &15.8 &12.5  &  &EC38/WMW7       \\
 67 &18 29 49.6 &01 14 57 &     &     &14.7  &* &EC40/GCNM22/\\
    &           &         &     &     &      &  &WMW48\\
 68 &18 29 49.6 &01 15 29 &     &     &14.6  &  &EC41/GCNM23/\\
    &           &         &     &     &      &  &WMW114\\
 69 &18 29 49.7 &01 13 49 &     &16.1 &14.1  &+ &EC42/GCNM25  \\
 70 &18 29 49.8 &01 12 33 &17.3 &16.0 &14.8  &! &EC43/GCNM26   \\
 71 &18 29 50.2 &01 12 17 &17.5 &16.5 &14.7  &  &EC45/GCNM29   \\
 72 &18 29 50.3 &01 12 41 &18.4 &17.3 &16.0  &! &K10           \\
 73 &18 29 50.6 &01 12 14 &18.5 &17.2 &16.1  &! &K11           \\
 74 &18 29 50.5 &01 08 58 &14.4 &13.0 &12.5  &  &WMW209        \\
 75 &18 29 50.5 &01 13 26 &     &     &17.3  &  &K12           \\
 76 &18 29 50.5 &01 12 23 &18.2 &17.1 &15.2  &  &EC47/GCNM32   \\
 77 &18 29 50.8 &01 16 15 &     &     &14.9  &+ &EC49/GCNM33   \\
 78 &18 29 51.2 &01 13 20 &15.8 &14.3 &13.4  &  &EC51/GCNM35/\\
    &           &         &     &     &      &  &WMW95\\
 79 &18 29 51.2 &01 12 41 &     &17.4 &15.2  &  &EC50/GCNM36   \\
 80 &18 29 51.2 &01 16 40 &15.9 &12.7 &10.9  &  &EC53/STGM27/\\
    &           &         &     &     &      &  &WMW24\\
 81 &18 29 51.4 &01 09 33 &15.7 &15.0 &      &  &WMW66\\
 82 &18 29 51.6 &01 13 14 &     &16.5 &15.5  &! &GCNM40/K13    \\
 83 &18 29 51.9 &01 14 42 &     &     &16.5  &  &K14           \\
 84 &18 29 51.9 &01 14 45 &     &     &15.6  &  &K15           \\
 85 &18 29 52.0 &01 13 22 &17.1 &14.7 &13.3  &  &EC56/GCNM44/K16\\
 86 &18 29 52.2 &01 15 48 &     &     &15.9  &  &SMM10-IR/WMW21 \\
 87 &18 29 52.2 &01 16 18 &17.9 &15.3 &13.8  &  &EC58/GCNM45/K17\\
 88 &18 29 52.1 &01 13 23 &19.4 &16.0 &14.5  &  &EC57/GCNM46/K18\\
 89 &18 29 52.4 &01 12 47 &19.0 &14.5 &12.5  &  &EC59/GCNM47   \\
 90 &18 29 52.5 &01 14 23 &     &     &14.5  &  &GCNM48/K19    \\
 91 &18 29 52.6 &01 14 45 &     &     &15.3  &  &K20           \\
 92 &18 29 52.7 &01 13 10 &     &15.6 &14.1  &  &EC61/GCNM50/K21\\
 93 &18 29 52.6 &01 13 26 &     &     &15.5  &  &EC60/GCNM51\\
 94 &18 29 52.5 &01 19 33 &16.8 &15.4 &14.3  &! &STGM33     \\
 95 &18 29 52.8 &01 14 36 &17.5 &16.1 &15.2  &! &GCNM52/K22 \\
 96 &18 29 52.8 &01 14 55 &     &     &14.1  &  &GCNM53     \\
\noalign{\smallskip}
\tableline
\end{tabular}
}
\end{center}
\end{table}

\setcounter{table}{0}

\begin{table}[!ht]
\caption{Continuation.}
\smallskip
\begin{center}
{\small
\begin{tabular}{rllrrrcl}
\tableline
\noalign{\smallskip}
No. &$\alpha_{2000}$&$\delta_{2000}$&$J$  &$H$  &$K$   &  &Name\\
\noalign{\smallskip}
\tableline
\noalign{\smallskip}
 97 &18 29 53.1 &01 06 18 &14.4 &13.1 &12.6  &  &WMW96      \\
 98 &18 29 53.1 &01 15 46 &     &     &14.8  &* &EC63/GCNM55       \\
 99 &18 29 53.1 &01 14 28 &     &16.9 &15.9  &! &GCNM56/K23         \\
100 &18 29 53.1 &01 12 28 &17.3 &15.4 &14.4  &! &EC64/GCNM57/\\
    &           &         &     &     &      &  &BD-Ser1/WMW86\\
101 &18 29 53.2 &01 14 38 &     &     &15.9  &  &K24              \\
102 &18 29 53.4 &01 14 23 &16.9 &14.9 &13.2  &+ &EC65/GCNM59        \\
103 &18 29 53.4 &01 17 01 &12.7 &11.3 &10.6  &  &EC67/GCNM60/STGM29/ \\
    &           &         &     &     &      &  &WMW81 \\
104 &18 29 53.4 &01 13 06 &14.9 &14.1 &13.5  &! &EC66/GCNM61/STGM14/ \\
    &           &         &     &     &      &  &WMW99  \\
105 &18 29 53.8 &01 12 47 &     &     &15.1  &  &WMW34       \\
106 &18 29 53.8 &01 13 31 &16.5 &14.2 &12.9  &  &EC68/GCNM63/WMW69    \\
107 &18 29 54.0 &01 07 11 &15.3 &14.2 &13.1  &  &STGM1/WMW39         \\
108 &18 29 54.0 &01 12 19 &     &     &16.4  &  &K25           \\
109 &18 29 54.0 &01 14 07 &19.7 &18.1 &16.7  &! &K26        \\
110 &18 29 54.0 &01 14 36 &18.7 &16.4 &15.8  &! &GCNM66/HCE165/K27  \\
111 &18 29 54.1 &01 14 43 &     &15.5 &15.3  &  &GCNM68/HCE166/K29   \\
112 &18 29 54.0 &01 20 19 &15.8 &14.5 &13.6  &  &STGM36     \\
113 &18 29 54.3 &01 12 19 &     &     &16.2  &  &K28          \\
114 &18 29 54.2 &01 16 36 &     &     &17.9  &  &K30           \\
115 &18 29 54.3 &01 15 03 &15.7 &12.6 &10.9  &  &EC69/GCNM69/CK10/ \\
    &           &         &     &     &      &  &STGM23 \\
116 &18 29 54.3 &01 18 17 &15.3 &14.8 &14.3  &! &STGM31     \\
117 &18 29 54.5 &01 14 48 &14.4 &13.3 &12.8  &  &EC70/GCNM70/WMW87\\
118 &18 29 54.5 &01 16 54 &     &     &14.2  &+ &EC71/GCNM71    \\
119 &18 29 55.1 &01 11 23 &16.5 &15.0 &14.0  &! &STGM6                   \\
120 &18 29 55.2 &01 13 23 &15.6 &13.2 &12.1  &  &EC73/GCNM75/STGM15/ \\
    &           &         &     &     &      &  &K2\_1/WMW62\\
121 &18 29 55.6 &01 08 34 &     &14.8 &14.3  &! & WMW56 \\
122 &18 29 55.3 &01 10 34 &14.0 &13.4 &12.9  &  & WMW201 \\
123 &18 29 55.6 &01 14 30 &15.0 &12.3 &10.7  &  &EC74/GCNM76/CK9/ \\
    &           &         &     &     &      &  &STGM21/K2\_3/WMW38\\
124 &18 29 55.9 &01 11 29 &18.8 &18.6 &18.0  &! &K31               \\
125 &18 29 55.8 &01 15 07 &18.3 &17.0 &14.2  &  &EC75/GCNM77         \\
126 &18 29 55.9 &01 15 11 &     &17.1 &14.5  &  &EC76/GCNM78     \\
127 &18 29 56.2 &01 10 58 &12.5 &11.7 &11.4  &  & WMW205 \\
128 &18 29 56.3 &01 12 18 &15.3 &14.2 &13.6  &! &EC77/GCNM80/STGM9/ \\
    &           &         &     &     &      &  &WMW204  \\
129 &18 29 56.1 &01 20 04 &16.6 &15.3 &14.3  &! &STGM35   \\
130 &18 29 56.4 &01 13 20 &     &     &17.3  &  &K32/WMW112              \\
131 &18 29 56.4 &01 15 32 &     &17.9 &15.0  &+ &GCNM83/HCE169      \\
132 &18 29 56.5 &01 13 01 &14.8 &12.5 &11.4  &  &EC79/GCNM84/STGM12/ \\
    &           &         &     &     &      &  &WMW80\\
133 &18 29 56.5 &01 12 41 &     &     &13.6  &  &EC80/GCNM85/WMW61         \\
134 &18 29 56.8 &01 13 31 &     &12.8 &12.2  &  &EC81                \\
135 &18 29 56.8 &01 14 46 &11.9 &10.3 &8.8   &  &SVS2/EC82/GCNM87/ \\
    &           &         &     &     &      &  &CK3/STGM22/WMW9\\
136 &18 29 56.7 &01 15 50 &     &17.3 &15.6  &+,! &GCNM88/HCE172  \\
137 &18 29 56.8 &01 16 09 &16.1 &15.7 &14.9  &! &EC83/GCNM89         \\
\noalign{\smallskip}
\tableline
\end{tabular}
}
\end{center}
\end{table}

\setcounter{table}{0}

\begin{table}[!ht]
\caption{Continuation.}
\smallskip
\begin{center}
{\small
\begin{tabular}{rllrrrcl}
\tableline
\noalign{\smallskip}
No. &$\alpha_{2000}$&$\delta_{2000}$&$J$  &$H$  &$K$   &  &Name\\
\noalign{\smallskip}
\tableline
\noalign{\smallskip}
138 &18 29 56.8 &01 12 49 &15.0 &12.4 &11.0  &  &EC84/GCNM90/STGM11/ \\
    &           &         &     &     &      &  &WMW85\\
139 &18 29 57.1 &01 11 29 &16.5 &15.2 &14.0  &  &STGM7          \\
140 &18 29 57.0 &01 15 41 &     &17.1 &15.7  &! &HCE173         \\
141 &18 29 57.1 &01 16 35 &16.0 &14.4 &13.6  &! &EC85/GCNM91         \\
142 &18 29 57.2 &01 12 59 &     &     &15.3  &  &GCNM92/HCE174        \\
143 &18 29 57.3 &01 14 50 &13.3 &11.5 &10.8  &  &EC86/GCNM93/WMW190  \\
144 &18 29 57.4 &01 13 50 &18.3 &15.5 &13.4  &  &EC87/GCNM94/K2\_4   \\
145 &18 29 57.5 &01 13 02 &     &     &11.7  &  &EC88/GCNM96/CK5   \\
146 &18 29 57.5 &01 13 06 &15.9 &13.5 &11.6  &  &EC89/GCNM97/STGM13    \\
147 &18 29 57.6 &01 14 05 &12.2 &9.3  &6.7   &  &SVS20S/EC90/GCNM98/\\
    &           &         &     &     &      &  &CK1/STGM18\\
148 &18 29 57.6 &01 14 07 &     &     &      &  &SVS20N        \\
149 &18 29 57.6 &01 10 47 & 9.2 &9.1  &9.1   &  &WMW193 \\
150 &18 29 57.7 &01 10 54 &8.1  &8.1  &8.1   &  &WMW192 \\
151 &18 29 57.7 &01 15 31 &15.8 &12.8 &10.8  &  &EC93/GCNM100/STGM25/ \\
    &           &         &     &     &      &  &CK12/WMW83 \\
152 &18 29 57.7 &01 12 28 &     &15.4 &12.6  &  &EC91/GCNM101/WMW70          \\
153 &18 29 57.8 &01 12 39 &     &14.1 &11.4  &  &EC94/GCNM102/WMW37        \\
154 &18 29 57.8 &01 12 47 &16.7 &12.3 & 9.8  &  &EC95/GCNM103  \\
155 &18 29 57.7 &01 12 52 &15.4 &12.0 &10.2  &  &EC92/GCNM104         \\
156 &18 29 58.2 &01 15 21 &13.0 &11.0 &9.8   &  &EC97/GCNM106/CK4/ \\
    &           &         &     &     &      &  &STGM24/WMW27\\
157 &18 29 58.1 &01 15 57 &     &     &16.1  &  &GCNM107/K33            \\
158 &18 29 58.3 &01 16 22 &     &     &15.3  &  &K34                \\
159 &18 29 58.4 &01 12 51 &     &14.6 &12.3  &  &EC98/GCNM110          \\
160 &18 29 58.5 &01 16 20 &     &     &11.0  &  &GCNM111               \\
161 &18 29 58.8 &01 14 25 &16.7 &14.2 &12.3  &  &EC103/GCNM112/STGM20/ \\
    &           &         &     &     &      &  &K2\_5/WMW4\\
162 &18 29 58.7 &01 16 21 &13.0 &11.4 &10.7  &  &EC100/GCNM113 \\
163 &18 29 58.7 &01 13 17 &     &     &13.4  &  &EC101    \\
164 &18 29 58.7 &01 13 04 &     &     &15.4  &  &EC102/GCNM114\\
165 &18 29 58.8 &01 15 03 &     &     &14.8  &  &EC104               \\
166 &18 29 59.0 &01 12 25 &     &     &14.6  &* &GCNM115/HCE175/WMW45    \\
167 &18 29 59.1 &01 14 41 &     &     &14.3  &  &EC106/GCNM116/K35     \\
168 &18 29 59.1 &01 11 20 &14.2 &12.6 &11.9  &  &WMW196\\
169 &18 29 59.1 &01 11 14 &     &     &14.4  &  &WMW198\\
170 &18 29 59.3 &01 08 11 &16.3 &14.5 &13.2  &  &STGM4              \\
171 &18 29 59.2 &01 16 27 &     &     &15.9  &+ &EC107/GCNM118          \\
172 &18 29 59.1 &01 14 09 &12.6 &10.5 &9.8   &  &EC105/GCNM119/CK8/ \\
    &           &         &     &     &      &  &STGM19/K2\_6\\
173 &18 29 59.2 &01 12 28 &     &     &15.3  &  &GCNM120               \\
174 &18 29 59.6 &01 11 57 &     &     &14.1  &  &K2\_7/HB1/WMW3              \\
175 &18 29 59.5 &01 15 54 &     &17.2 &14.9  &+ &GCNM123/HCE176          \\
176 &18 29 59.5 &01 14 53 &     &17.2 &15.3  &  &EC109/GCNM124         \\
177 &18 30 00.0 &01 14 04 &15.6 &14.0 &12.8  &  &EC114/GCNM131/STGM17/\\
    &           &         &     &     &      &  &WMW28\\
178 &18 30 00.1 &01 03 06 &14.2 &12.7 &12.1  &  &WMW156\\
179 &18 30 00.2 &01 14 45 &     &16.3 &15.0  &! &EC116/GCNM132/K36         \\
180 &18 30 00.5 &01 13 40 &13.3 &11.1 &10.1  &  &EC117/GCNM135/CK6/\\
    &           &         &     &     &      &  &WMW216  \\
\noalign{\smallskip}
\tableline
\end{tabular}
}
\end{center}
\end{table}

\setcounter{table}{0}

\begin{table}[!ht]
\caption{Continuation.}
\smallskip
\begin{center}
{\small
\begin{tabular}{rllrrrcl}
\tableline
\noalign{\smallskip}
No. &$\alpha_{2000}$&$\delta_{2000}$&$J$  &$H$  &$K$   &  &Name\\
\noalign{\smallskip}
\tableline
\noalign{\smallskip}
181 &18 30 00.5 &01 15 19 &18.0 &12.6 & 8.9  &  &EC118//GCNM136/CK2/\\
    &           &         &     &     &      &  &WMW158     \\
182 &18 30 00.9 &01 16 51 &16.4 &15.1 &14.2  &! &EC120/GCNM141           \\
183 &18 30 00.9 &01 22 04 &15.9 &15.1 &14.5  &! &WMW77\\
184 &18 30 01.0 &01 13 24 &     &15.3 &12.8  &  &EC121/GCNM142/WMW30          \\
185 &18 30 01.1 &01 15 05 &15.5 &13.2 &12.0  &  &EC122/GCNM146/CK12/\\
    &           &         &     &     &      &  &WMW195 \\
186 &18 30 01.5 &01 15 15 &15.7 &15.0 &15.1  &  &EC123/GCNM148/K37/\\
    &           &         &     &     &      &  &SVS1/WMW31   \\
187 &18 30 01.5 &01 20 36 &15.8 &14.9 &14.2  &! &STGM37         \\
188 &18 30 01.7 &01 04 43 &16.1 &14.2 &13.1  &  & WMW92 \\
189 &18 30 01.9 &01 14 01 &16.0 &14.6 &12.6  &  &EC125/GCNM154/CK7/\\
    &           &         &     &     &      &  &STGM16/WMW1\\
190 &18 30 02.4 &01 12 45 &     &     &15.5  &+ &EC128/GCNM156/WMW47        \\
191 &18 30 02.4 &01 14 15 &     &     &17.0  &+ &EC127/GCNM157         \\
192 &18 30 02.7 &01 12 30 &15.2 &11.8 &10.0  &  &EC129/GCNM160/STGM10/\\
    &           &         &     &     &      &  &WMW10   \\
193 &18 30 02.8 &01 16 16 &     &     &15.4  &+ &EC132/GCNM162          \\
194 &18 30 02.8 &01 13 23 &15.7 &15.3 &14.7  &! &EC131/GCNM163          \\
195 &18 30 03.0 &01 18 11 &14.0 &13.2 &12.6  &  &STGM30                 \\
196 &18 30 03.0 &01 19 50 &16.3 &15.9 &15.1  &! &WMW189\\
197 &18 30 03.3 &01 16 20 &12.6&10.8&10.3&&EC135/GCNM165/STGM26/\\
    &           &         &     &     &      &  &WMW78\\
198 &18 30 03.5 &01 16 15 &14.3 &13.4 &13.0  &! &EC138      \\
199 &18 30 03.6 &01 16 41 &     &     &14.2  &  & WMW174 \\
200 &18 30 03.6 &01 15 04 &     &     &16.4  &  &K38    \\
201 &18 30 03.6 &01 12 41 &16.2 &13.4 &12.2  &* &EC141     \\
202 &18 30 03.7 &01 16 10 &     &15.7 &13.6  &  &EC142      \\
203 &18 30 04.1 &01 06 56 &16.8 &14.9 &14.2  &! & WMW60 \\
204 &18 30 04.9 &01 12 12 &     &     &16.1  &  &K39       \\
205 &18 30 04.8 &01 14 41 &13.7 &12.8 &12.3  &  &EC149/WMW98       \\
206 &18 30 04.9 &01 12 37 &16.0 &14.3 &13.6  &! &EC152/K40/WMW54   \\
207 &18 30 05.5 &01 14 25 &15.2 &14.4 &14.1  &! & WMW203 \\
208 &18 30 05.7 &01 13 50 &17.0 &13.9 &12.2  &  &EC160       \\
209 &18 30 05.9 &01 13 46 &     &14.0 &12.5  &  &EC161       \\
210 &18 30 05.9 &01 21 42 &14.8 &14.0 &13.3  &  &STGM38      \\
211 &18 30 06.1 &01 06 17 &12.9 &11.8 &11.2  &  &WMW84\\
212 &18 30 06.9 &01 18 24 &14.8 &13.7 &12.9  &  &STGM32     \\
213 &18 30 07.7 &01 12 05 &12.2 &10.9 &10.0  &  &STGM8/WMW73       \\
214 &18 30 08.3 &01 11 39 &19.2 &17.9 &16.9  &! &K42         \\
215 &18 30 08.2 &01 12 41 &     &     &14.4  &  &K41         \\
216 &18 30 08.6 &01 17 01 &17.1 &14.9 &13.2  &  &STGM28      \\
217 &18 30 08.8 &01 12 02 &16.1 &14.6 &13.7  &! &K43        \\
218 &18 30 08.9 &01 15 47 &     &     &14.3  &  &K44        \\
219 &18 30 09.2 &01 17 53 &15.4 &14.0 &13.0  &  & WMW67\\
220 &18 30 09.3 &01 14 07 &18.4 &16.6 &15.2  &! &K45         \\
221 &18 30 09.4 &01 02 47 &10.6 & 8.9 & 8.1  &  &WMW170\\
222 &18 30 09.7 &01 15 21 &18.6 &17.9 &16.8  &! &K46        \\
223 &18 30 10.6 &01 16 21 &19.2 &18.5 &17.1  &! &K47        \\
224 &18 30 11.2 &01 12 12 &16.8 &15.2 &14.2  &! &K48        \\
225 &18 30 11.1 &01 12 57 &17.5 &16.3 &15.2  &! &K49        \\
\noalign{\smallskip}
\tableline
\end{tabular}
}
\end{center}
\end{table}

\setcounter{table}{0}

\begin{table}[!ht]
\caption{Continuation.}
\smallskip
\begin{center}
{\small
\begin{tabular}{rllrrrcl}
\tableline
\noalign{\smallskip}
No. &$\alpha_{2000}$&$\delta_{2000}$&$J$  &$H$  &$K$   &  &Name\\
\noalign{\smallskip}
\tableline
\noalign{\smallskip}
226 &18 30 11.1 &01 12 38 &     &12.5 &12.0  &  &WMW101\\
227 &18 30 11.6 &01 13 57 &     &     &14.4  &  &K50        \\
228 &18 30 12.0 &01 16 45 &     &     & 8.7  &  &K51/SVS6\\
229 &18 30 12.3 &01 17 45 &     &     &15.7  &  &K52\\
230 &18 30 12.5 &01 15 33 &     &     &15.1  &  &K53\\
231 &18 30 13.3 &01 02 49 &13.7 &12.8 &12.4  &  &WMW74\\
232 &18 30 13.3 &01 15 31 &     &     &15.8  &  &K54\\
233 &18 30 13.5 &01 12 18 &     &     &16.4  &  &K55\\
234 &18 30 13.3 &01 17 46 &     &     &16.8  &  &K3\_11\\
235 &18 30 14.0 &01 08 52 &13.8 &12.7 &12.4  &  &WMW166 \\
236 &18 30 14.4 &01 04 06 &15.3 &14.3 & 13.9 &! &WMW89 \\
237 &18 30 15.4 &01 17 30 &     &     &15.8  &  &K56\\
238 &18 30 15.6 &01 17 31 &     &     &14.6  &  &K57\\
239 &18 30 16.8 &01 19 17 &16.2 &15.8 &15.2  &! &WMW187\\
240 &18 30 17.0 &01 13 08 &16.5 &15.2 &14.2  &! &WMW76  \\
241 &18 30 17.3 &01 21 33 & 9.7 & 8.2 & 7.5  &  &WMW161 \\
242 &18 30 18.2 &01 14 17 &13.3 &11.8 &10.9  &  &WMW40   \\
243 &18 30 22.4 &01 20 44 &13.1 &12.2 &11.9  &  &WMW157   \\
244 &18 30 22.7 &01 13 24 &16.0 &15.0 &14.6  &! &WMW58     \\
245 &18 30 22.8 &01 16 21 &15.6 &14,4 &13.6  &! &WMW55     \\
246 &18 30 23.1 &01 20 10 &13.3 &12.6 &12.2  &  &WMW128    \\
247 &18 30 23.4 &01 05 05 &13.4 &12.4 &11.8  &  &WMW33     \\
248 &18 30 24.4 &01 19 51 &12.4 &11.9 &11.8  &  &WMW220     \\
249 &18 30 27.7 &01 22 35 &15.9 &15.3 &14.5  &! &WMW102    \\
250 &18 30 37.3 &01 16 09 &10.4 & 8.8 & 8.2  &  &WMW140    \\
251 &18 30 41.2 &01 13 16 &14.5 &13.7 &13.2  &! &WMW88     \\
252 &18 30 44.5 &01 23 30 &     &14.7 &14.5  &! &WMW90     \\
253 &18 30 53.7 &01 11 21 &11.1 & 9.6 & 8.9  &  &WMW143    \\
\noalign{\smallskip}
\tableline
\end{tabular}
}
\end{center}
(\$) Labels correspond to the following references: ESO Ha 279a: Aspin et al.
1994; EC: Eiroa \& Casali 1992; GCNM: Giovannetti et al. 1998; STGM: Sogawa
et al. 1997;  K: Kaas 1999; SMM9, SMM10: Hodapp 1999; DEOS: Hodapp et al 1996;
BD-Ser,  KCML: Klotz et al. 2004; HCE: Horrobin et al. 1997; CK: Churchwell
\& Koorneef 1986; SVS: Strom et al. 1976, HB: Hurt \& Barsony, 1996;
WMW: Winston et al. (2007) (\#)
Recent spectra show that this object is in fact a background M-type giant
star (Aspin \& Greene, 2007).
(*) Identified using ISO data (Kaas et al. 2004). (+) Serpens object candidate
suggested in this work based on $K$-brightness variability ($\Delta K > 0.5 $
 mag. (!) Brown dwarf candidates suggested in this work.
\end{table}

\begin{landscape}
\begin{table}[!ht]
\caption{Cross-correlation of the near-IR source names with  other
wavelength names.}
\smallskip
{\small
\begin{tabular}{rlllrrrcrl}
\tableline
\noalign{\smallskip}
No. &Optical& Near-IR          &ISO  & Spitzer   &Submm&VLA  &XMM & Chandra &  Class   \\
\noalign{\smallskip}
\tableline
\noalign{\smallskip}
 11 & &ESO H$\alpha$279a/WMW36                & 159 & 104 &       &      &     9         &    & flat\\
 12 & &ESO H$\alpha$279b                      & 160 &     &       &      &               &    & II\\
 24 & &STGM3/WMW103                           & 207 &     &       &      &               &    &  II \\
 28 & &KCML6/WMW215                           &     &     &       &      &               & 21 & III  \\
 31 & &STGM2/WMW65                            & 219 & 125 &       &      &               &    & II \\
 34 & &EC11/WMW100                            & 224 & 128 &       &      &               & 23 & II  \\
 36 & &EC13                                   & 226 &     &       &      &               &    & II     \\
 39 & &EC21                                   & 231 &     &       &      &               &    & II     \\
 44 & &EC23                                   & 232 &     &       &      &               &    & II    \\
 48 & &EC26                                   & 234 &     &       &      &               &    & flat  \\
 49 & &EC28/WMW5                              & 237 & 131 &       &      &               &    & I,flat   \\
 57 & &SMM9-IR/WMW23                          & 241 & 135 & SMM9  &VLA5  &               &    & 0,I  \\
 60 & &EC33/GCNM13/K8/                        & 242 &     &       &      &               &    & II\\
    & &KCML7/WMW106                           &     &     &       &      &               &    &   \\
 63 & &DEOS/K4\_5/WMW11                       & 250 & 137 &       &      &               &    & I   \\
 64 & &EC37/GCNM19/                           & 249 & 138 &       &      &               &    & I         \\
    & &WMW6                                   &     &     &       &      &               &    &   \\
 66 & &EC38/WMW7                              & 254 & 139 &       &      &               & 28 & I    \\
 67 & &EC40/GCNM22/                           & 253 & 142 &       &      &               &    & I     \\
    & &WMW48                                  &     &     &       &      &               &    &   \\
 68 & &EC41/GCNM23/                           & 258a&     & SMM1  &VLA7  &               &    & 0,I  \\
    & &WMW114                                 &     &     &       &      &               &    &   \\
 78 & &EC51/GCNM35/                           & 266 & 147 &       &      &               & 31 & II   \\
    & &WMW95                                  &     &     &       &      &               &    &   \\
 80 & &EC53/STGM27/                           & 265 &     & SMM5  & VLA9 &182951.2+011640& 30 & I\\
    & &WMW24                                  &     &     &       &      &               &    &   \\
 85 & &EC56/GCNM44/K16                        & 266 &     &       &      &               &    & II \\
 86 & &SMM10-IR/WMW21                         & 270 & 150 & SMM10 &VLA10 &               &    & 0,I \\
 89 & &EC59/GCNM47                            & 272 &     &       &      &               &    & II  \\
 96 & &GCNM53                                 & 276 &     &       &VLA11 &               & 33 & I    \\
\noalign{\smallskip}
\tableline
\end{tabular}
}
\end{table}
\end{landscape}
\setcounter{table}{1}
\begin{landscape}
\begin{table}[!ht]
\caption{Continuation.}
\smallskip
{\small
\begin{tabular}{rlllrrrcrl}
\tableline
\noalign{\smallskip}
No. &Optical& Near-IR&ISO& Spitzer&Submm  &VLA  &XMM& Chandra&  Class\\
\noalign{\smallskip}
\tableline
\noalign{\smallskip}
 98 & &EC63/GCNM55                            & 277 &     &       &      &               &    & I     \\
100 & &EC64/GCNM57/                           &     & 160 &       &      &               &    & II    \\
    & &BD-Ser1/WMW86                          &     &     &       &      &               &    &   \\
103 & &EC67/GCNM60/                           & 283 & 162 &       &      &182953.7+011702& 34 & II \\
    & &STGM29/WMW81                           &     &     &       &      &               &    &   \\
104 & &EC66/GCNM61/                           & 279 &     &       &      &               &    & II \\
    & &STGM14/WMW99                           &     &     &       &      &               &    &   \\
106 & &EC68/GCNM63/                           & 285 & 163 &       &      &               &    & II  \\
    & &WMW69                                  &     &     &       &      &               &    &   \\
107 & &STGM1/WMW39                            &     & 164 &       &      &               &    & flat   \\
115 & &EC69/GCNM69/                           & 289 &     &       &      &               &    & II \\
    & &CK10/STGM23                            &     &     &       &      &               &    &   \\
117 &GEL2&EC70/GCNM70/                        & 291 &     &       &      &               & 35 &  II \\
    & &WMW87                                  &     &     &       &      &               &    &   \\
120 &GEL3&EC73/GCNM75/K2\_1/                  & 294 & 168 &       &      &               &    & flat\\
    & &STGM15/WMW62                           &     &     &       &      &               &    &   \\
123 &GEL4&EC74/GCNM76/CK9/                    & 298 & 171 &       &      &182955.7+011428& 37 & II \\
    & &STGM21/K2\_3/WMW38                     &     &     &       &      &               &    &   \\
128 & &EC77/GCNM80/                           &     &     &       &      &               & 39 & III\\
    & &STGM9/WMW204                           &     &     &       &      &               &    &   \\
132 &GEL5&EC79/GCNM84/                        & 304 & 174 &       & VLA12&               & 40 & II\\
    & &STGM12/WMW80                           &     &     &       &      &               &    &   \\
133 && EC80/GCNM85/                           & 306 & 175 &       &      &               & 42 & I \\
    & &WMW61                                  &     &     &       &      &               &    &   \\
135 &GEL6&SVS2/EC82/CK3/                      & 307 & 176 &       &      &               & 43 & flat\\
    & &GCNM87/STGM22/WMW9                     &     &     &       &      &               &    &   \\
138 &GEL7&EC84/GCNM90/                        & 309 &     &       & VLA14&               & 44 & II\\
    & &STGM11/WMW85                           &     &     &       &      &               &    &   \\
\noalign{\smallskip}
\tableline
\end{tabular}
}
\end{table}
\end{landscape}
\setcounter{table}{1}
\begin{landscape}
\begin{table}[!ht]
\caption{Continuation.}
\smallskip
{\small
\begin{tabular}{rlllrrrcrl}
\tableline
\noalign{\smallskip}
No. &Optical& Near-IR&ISO& Spitzer&Submm  &VLA  &XMM& Chandra&  Class\\
\noalign{\smallskip}
\tableline
\noalign{\smallskip}
143 &GEL8&EC86/GCNM93/                        &     &     &       &      &182757.5+011448& 45 & III \\
    & &WMW190                                 &     &     &       &      &               &    &   \\
144 & &EC87/GCNM94/                           & 313 &     &       &      &               &    & I \\
    & &/K2\_4                                 &     &     &       &      &               &    &   \\
145 &GEL9&EC88/GCNM96/                        & 312 & 179 &       & VLA15&               & 46 & I \\
    & &CK5                                    &     &     &       &      &               &    &   \\
146 &&EC89/GCNM97/                            & 312 & 181 &       &      &               &    &   \\
    & &STGM13                                 &     &     &       &      &               &    &   \\
147 &GEL10&SVS20S/EC90/CK1                    & 314 & 182 & SMM6  & VLA16&182957.7+011407& 48 & flat\\
    & &GCNM98/STGM18/WMW35                    &     &     &       &      &               &    &   \\
148 &&SVS20N                                  & 314 &     &       &      &               &    & flat      \\
151 &&EC93/GCNM100/                           & 319 & 183 &       &      &182957.9+011530& 50 & II \\
    & &STGM25/CK12/WMW83                      &     &     &       &      &               &    &   \\
152 &&EC91/GCNM101/                           & 320 & 184 &       &      &               &    & flat   \\
    & &WMW70                                  &     &     &       &      &               &    &   \\
153 &&EC94/GCNM102/                           & 318 & 186 &       &      &               & 52 & flat   \\
    & &WMW37                                  &     &     &       &      &               &    &   \\
154 &&EC95/GCNM103                            & 317 & 187 &       & VLA17&182957.9+011246& 53 & II\\
155 &&EC92/GCNM104                            & 317 & 185 &       &VLA17'&               &    & I \\
156 &GEL12& EC97/GCNM106/                     & 321 & 188 &       &      &182958.4+011520& 55 & II\\
    & &CK4/STGM24/WMW27                       &     &     &       &      &               &    &   \\
159 &&EC98/GCNM110                            & 322 &     &       &      &               & 56 & flat     \\
161 &&EC103/GCNM112/                          & 326 & 190 &       &      &               &    & I\\
    & &STGM20/K2\_5/WMW4                      &     &     &       &      &               &    &   \\
166 &      & GCNM115/HCE175/                      & 327 & 192 & &       &               &    & I    \\
    &      & WMW46                                &     &     &       &      &               &    &   \\
172 &GEL13 & EC105/GCNM119/                       & 328 & 194 & &       &182959.3+011408& 59 & II \\
    &      & CK8/STGM19/K2\_6/WMW59               &     &     &       &      &               &    &   \\
174 &      & K2\_7/HB1/WMW3                       & 330 & 197 & &       & +             & 60 & I             \\
\noalign{\smallskip}
\tableline
\end{tabular}
}
\end{table}
\end{landscape}
\setcounter{table}{1}
\begin{landscape}
\begin{table}[!ht]
\caption{Continuation.}
\smallskip
{\small
\begin{tabular}{rlllrrrcrl}
\tableline
\noalign{\smallskip}
No. &Optical& Near-IR&ISO& Spitzer&Submm  &VLA  &XMM& Chandra&  Class\\
\noalign{\smallskip}
\tableline
\noalign{\smallskip}
177 &GEL14 & EC114/GCNM131/                       &     & 200 & &       &               & 62 & I/II\\
    &      &STGM17/WMW28                          &     &     &       &      &               &    &   \\
180 &GEL15 &EC117/GCNM135/                        & 338 &     & & VLA20 &183000.7+011338& 65 & II  \\
    &      &CK6/WMW216                            &     &     &       &      &               &    &   \\
184 &      &EC121/GCNM142/                        & 341 & 204 & &       &               & 66 & flat        \\
    &      &WMW30                                 &     &     &       &      &               &    &   \\
185 &GEL16 &EC122/GCNM146/                        &     &     & &       &183001.3+011500& 67 & III\\
    &      &CK12/WMW195                           &     &     &       &      &               &    &   \\
189 &      &EC125/GCNM154/                        & 345 & 207 & &       &               &    & flat\\
    &      &CK7/STGM16/WMW1                       &     &     &       &      &               &    &   \\
192 &      &EC129/GCNM160/                        & 347 & 208 & &       &               &    & flat  \\
    &      &STGM10/WMW10                          &     &     &       &      &               &    &   \\
197 &GEL18 &EC135/GCNM165/                        & 348 & 209 & &       &183003.5+011620& 69 & II\\
    &GGD29 &STGM26/WMW78                          &     &     &       &      &               &    &   \\
201 &      &EC141                                 & 351 &     & &       &               &    & II    \\
205 &      &EC149/WMW98                           &     & 215 & &       &               & 71 & II\\
206 &      &EC152/K40/WMW54                       & 356 & 216 & &       &               &    & II  \\
213 &      &STGM8/WMW73                           & 366 & 220 & &       &183007.7+011204& 75 & II      \\
\noalign{\smallskip}
\tableline
\end{tabular}
}

(!)  Labels  in the  different  columns  correspond  to the  following
references:  column 1:  numbers of  the  near-IR objects  in Table  1;
column 2: G\'omez de Castro et  al. (1988); column 3: references as in
Table  1;  column  4:  Kaas   et  al.  (2004);  column  5:  Harvey  et
al.  (2007a); column 6:  Casali et  al. (1993),  Davis et  al. (1999);
column 7: Eiroa et al. (2005); column 8: Preibisch (2003,2004); column
9: Giardino et al. (2007);  column 10: evolutionary classes taken from
Kaas et al. (2004), Eiroa et al. (2005), and Winston et al. (2007).
\end{table}
\end{landscape}


\begin{thebibliography}{}
\bibitem[n(0)]{} Alexander, R. D., Casali, M. M., Andre P., Persi, P., \& Eiroa, C.
2003, \aap, 401, 613
\bibitem[n(0)]{} Aspin, C. \& Greene, T.P. 2007, \aj, 133, 568
\bibitem[n(0)]{} Aspin, C., Reipurth, B., \& Lehmann, T. 1994, \aap, 288, 165
\bibitem[n(0)]{} Baraffe, I., Chabrier, G., Barman, T.S., Allard, F.,
\& Hauschildt, P.H. 2003, \aap, 402, 701
\bibitem[n(0)]{} Bally, J. \& Lada, C.J. 1983, \apj, 265, 824
\bibitem[n(0)]{} Bernes, C. 1977, \aaps, 29, 65
\bibitem[n(0)]{} Blair, G.N., Davis, J.H., \& Dickinson, D.F. 1978, \apj, 226, 435
\bibitem[n(0)]{} Blake, G.A. \& Boogert, A.C.A. 2004, \apj,  606, L73
\bibitem[n(0)]{} Boogert, A.C.A., Pontoppidan, K.M., Knez, A., et al. 2008
\aj, 678, 985
\bibitem[n(0)]{} Brown, D.W., Chandler, C.J., Carlstrom, J.E., et al. 2000, \mnras
~319, 154
\bibitem[n(0)]{} Cambr\'esy, L. 1999, \aap, 345, 965
\bibitem[n(0)]{} Caratti o Garatti, A., Giannini, T., Nisini, B., \& Lorenzetti,
D. 2006, \aap, 449, 1077
\bibitem[n(0)]{} Casali, M.M. \& Eiroa, C. 1996a, \aap, 306, 427
\bibitem[n(0)]{}  Casali, M.M.  \& Eiroa,  C. 1996b,  in {\it  Cool stars;
stellar systems;  and the Sun}, eds.  R. Pallavicini \&  A. K. Dupree,
ASP Conference Series, volume 109, p. 713
\bibitem[n(0)]{} Casali, M.M., Eiroa, C., \& Duncan, W. D. 1993, \aap, 275, 195
\bibitem[n(0)]{} Chavarr\'\i a-K, C., de Lara, E., Finkenzeller, U., Mendoza,
E.E., \& Ocegueda, J. 1988, \aap, 197, 151
\bibitem[n(0)]{} Chiar, J. E., Adamson, A. J., Kerr, T. H., \& Whittet, D. C. B.
1994, \apj, 426, 240
\bibitem[n(0)]{} Chini, R., Reipurth, B., Sievers, A., et al. 1997, \aap, 325, 542
\bibitem[n(0)]{} Chini, R., Ward-Thompson, D., Kirk, J. M., et al. 2001, \aap, 369,
155
\bibitem[n(0)]{} Ciardi, D.R., Telesco, C. M., Packham, C., et al. 2005, \apj, 629
897
\bibitem[n(0)]{} Clark, F.O. 1990, priv. comm.
\bibitem[n(0)]{} Clark, F.O. 1991, 	\apjs, 75, 611
\bibitem[n(0)]{} Clark, F.O. \& Turner, B.E. 1987, \aap, 176, 114
\bibitem[n(0)]{} Cohen, M. \& Kuhi, L.V. 1979, \apjs, 41, 743
\bibitem[n(0)]{} Connelley, M.S., Reipurth, B., \& Tokunaga, A.T. 2007, \aj, 133, 1528
\bibitem[n(0)]{} Covey, R., Greene, T.P., Doppmann, G.W., \& Lada, C.J. 2006,
\aj, 131, 512
\bibitem[n(0)]{} Curiel, S., Rodr\'\i guez, L.F., G\'omez, J.F., et al. 1996, \apj, 456, 677
\bibitem[n(0)]{} Curiel, S., Rodr\'\i guez, L.F., Moran, J.M., \& Cant\'o, J. 1993, \apj, 415, 191
\bibitem[n(0)]{} Dame T.M., Hartmann, D., \& Thaddeus, P. 1987, \apj, 547, 792
\bibitem[n(0)]{} Dame T.M. \& Thaddeus, P. 1985, \apj, 297, 751
\bibitem[n(0)]{} Dame T.M., Ungerechts, H., Cohen, R.S. et al. 2001, \apj, 322, 706
\bibitem[n(0)]{} Davis, C.J., Matthews, H.E., Ray, T.P., Dent, W.R.F., \& Richer,
J.S. 1999, \mnras, 309, 141
\bibitem[n(0)]{} de Lara, E. \& Chavarria-K, C. 1989, Rev. Mex. Astron. Astrofis.
18, 180
\bibitem[n(0)]{} de Lara, E., Chavarria-K, C., \& L\'{o}pez-Molina, G. 1991, \aap, 243, 139
\bibitem[n(0)]{} Dinger, A.S.C. \& Dickinson, D.F. 1980, \aj, 85, 1247
\bibitem[n(0)]{} Djupvik, A.A., Andr\'e, Ph., Bontemps, S., et al. 2006,
\aap, 458, 789
\bibitem[n(0)]{} Dobashi, K., Uehara, H., Kandori, R., et al. 2005, PASJ, 57, No.
SP1, S1
\bibitem[n(0)]{} Doppmann, G.W., Greene, T.P., Covet, K.R., \& Lada, C.J. 2005,
\aj, 130, 1145
\bibitem[n(0)]{} Dorschner, J. \& G\"urtler, J. 1963, AN, 287, 257
\bibitem[n(0)]{} Duch\^ene, G., Bontemps, S., Bouvier, J., et al. 2007, \aap, 476,
229
\bibitem[n(0)]{} Eiroa, C. 1991, in {\it Low Mass Star Formation in Southern Molecular
Clouds}, ed. Bo Reipurth, ESO Scientific Report N. 11, p. 197
\bibitem[n(0)]{} Eiroa, C. \& Casali, M.M. 1989, \aap, 223, L17
\bibitem[n(0)]{} Eiroa, C. \& Casali, M.M. 1992, \aap, 262, 468
\bibitem[n(0)]{} Eiroa, C., Mora, A., \& Djupvik, A.A. 2008, in preparation
\bibitem[n(0)]{} Eiroa, C., Djupvik, A.A., \& Casali, M.M. 2006, AN, 327, 14
\bibitem[n(0)]{} Eiroa, C. \& Hodapp, K.-W 1989, \aap, 210, 345
\bibitem[n(0)]{} Eiroa, C. \& Leinert, C. 1987, \aap, 188, 46
\bibitem[n(0)]{} Eiroa, C., Lenzen, R., Leinert, Ch., \& Hodapp, K.-W
1987, \aap, 179, 171
\bibitem[n(0)]{} Eiroa, C.; Palacios, J., \& Casali, M. M. 1997a, in {\it Planets Beyond
the Solar System and the Next Generation of Space Missions}. ed. D.
Soderblom, ASP Conference Series, Vol. 119, p.107
\bibitem[n(0)]{} Eiroa, C., Palacios, J., Eisl\"offel, J., Casali, M.M.,
\& Curiel, S. 1997b, in {\it Herbig-Haro Flows and the Birth of Stars}, eds.
B. Reipurth and C. Bertout, IAU Symposium 182, p. 103
\bibitem[n(0)]{} Eiroa, C., Torrelles, J.M., Curiel, S., \& Djupvik, A.A. 2005, \aj, 130, 643
\bibitem[n(0)]{} Eiroa, C., Torrelles, J.M., G\'omez, J.F., et al. 1992, PASJ, 44,
155
\bibitem[n(0)]{} Eisner, J.A., Lane, B.F., Akeson, R.L., Hildebrand, L.A., \& Sargent,
A.I., 2003, \apj,  588, 360
\bibitem[n(0)]{} Eisner, J.A., Lane, B.F., Hildebrand, L.A., Akeson, R.L., \& Sargent,
A.I, 2004, \apj,  613, 1049
\bibitem[n(0)]{} Enoch, M.L., Glenn, J., Evans II, N.J., et al. 2007, \apj, 666, 982
\bibitem[n(0)]{} Evans II, N.J., Allen, L.E., Blake, G.A., et al. 2003, \pasp
~115, 965
\bibitem[n(0)]{} Festin, L. 1998, \aap, 336, 883
\bibitem[n(0)]{} Flaherty, K.M., Pipher, J.L., Megeath, S.T., et al. 2007, \apj, 663, 1069
\bibitem[n(0)]{} Forbrich, J., Mass, M., Ros, E., Brunthaler, A., \& Menten, K.M.
2007, \aap, 985, 992
\bibitem[n(0)]{} Fuente, A., Ceccarelli, C., Neri, R., et al. 2007, \aap, 468, L37
\bibitem[n(0)]{} Furuya, R.S., Kitamura, Y., Wootten, A., Claussen, M.J.,
\& Kawabe, R. 2003, \apjs, 144, 71
\bibitem[n(0)]{} Geers, V.C., Augereau, J.-C., Pontoppidan, K.M., et al. 2006,
\aap, 459, 545
\bibitem[n(0)]{} Geers, V.C., van Dishoeck, E.F., Visser, R., et al. 2007
\aap, 476, 279
\bibitem[n(0)]{} Giardino, G., Favata, F., Micela, G., Sciortino, S., \& Winston,
E. 2007, \aap, 275, 288
\bibitem[n(0)]{} Giovannetti, P., Caux, E., Nadeau, D., \& Monin, J.-L. 1998, \aap, 330, 990
\bibitem[n(0)]{} G\'omez de Castro, A.I 1997, \aap, 323, 541
\bibitem[n(0)]{} G\'omez de Castro, A.I., Eiroa, C., \& Lenzen, R. 1988, \aap, 201,
299
\bibitem[n(0)]{} Gregersen, E.M., Evans II, N.J., Zhou, S., \& Choi, M. 1997, \apj
~484, 256
\bibitem[n(0)]{} Gutermuth, R.A., Bourke, T.L., Allen, L.E., et al. 2008, \apj, 673, L151
\bibitem[n(0)]{} Gyulbudaghian, A.L., Glushkov, Yu. I., \& Denisyuk, E.K. 1978,
\apj, 224, L137
\bibitem[n(0)]{} Haisch, K.E. Jr., Barsony, M., Greene, T.P., \& Ressler, M.E.
2002, \aj, 124, 2841
\bibitem[n(0)]{} Haisch, K.E. Jr., Barsony, M., Ressler, M.E, \& Greene, T.P. 2006,
\aj, 132, 2675
\bibitem[n(0)]{} Haisch, K.E. Jr., Greene, T.P.,  Barsony, M., \& Stahler S.W. 2004,
\aj, 127, 1747
\bibitem[n(0)]{} Hartigan, P. \& Lada, C. J. 1985, \apjs, 59, 383
\bibitem[n(0)]{} Harvey, P.M., Chapman, N., Lai, S.P.,  et al. 2006, \apj, 644, 307
\bibitem[n(0)]{} Harvey, P.M., Mer\'\i n, B., \& Huard, L.T., et al. 2007a,  \apj,  663, 1149
\bibitem[n(0)]{} Harvey, P.M., Rebull, L.M., \& Brooke, T., et al. 2007b, \apj, 663, 1139
\bibitem[n(0)]{} Harvey, P.M., Wilking, B.A., \& Joy, M. 1984, \apj, 278, 156
\bibitem[n(0)]{} Herbst, T.M., Beckwith, S.V.W., \& Robberto, M. 1997, \apj, 486, L59
\bibitem[n(0)]{} Herbst, W. \& Shevchenko, V.S. 1999, \aj, 118, 1034
\bibitem[n(0)]{} Ho, P.T.P. \& Barrett, A.H. 1980, \apj, 237, 38
\bibitem[n(0)]{} Hodapp, K.-W. 1999, \aj, 118, 1338
\bibitem[n(0)]{} Hodapp, K.-W., Hora, J.L., Raymer, J.T., Pickles, A.J., \& Ladd, E.F.
1996, \apj, 468, 861
\bibitem[n(0)]{} Hogerheijde, M.R., van Dishoeck, E.W., Salverda, J. M., \& Blake, G.
A. 1999, \apj, 513, 350
\bibitem[n(0)]{} Horrobin, M. J., Casali, M. M., \& Eiroa, C. 1997, \aap, 320, L41
\bibitem[n(0)]{} Huard, T. L., Weintraub, D. A., \& Kastner, J. H. 1997, \mnras, 290,
598
\bibitem[n(0)]{} Hurt, R.L. \& Barsony, M. 1996, \apj, 460, L45
\bibitem[n(0)]{} Hurt, R.L., Barsony, M., \& Wootten, A.H. 1996, \apj, 456, 686
\bibitem[n(0)]{} Isella, A., Testi, L., \& Natta, A. 2006, \aap, 451, 951
\bibitem[n(0)]{} Kaas, A.A. 1999, \aj, 118, 558
\bibitem[n(0)]{} Kaas, A.A., Olofsson, G., Bontemps S., et al. 2004, \aap, 421,
623
\bibitem[n(0)]{} Kessler-Silacci, J., Augereau, J.-Ch., Dullemond, C.P., et al.
2006, \apj, 639, 275
\bibitem[n(0)]{} King, D.J., Scarrott, S.M., \& Taylor, K.N.R. 1983, \mnras, 202, 1087
\bibitem[n(0)]{} Klotz, A., Caux, E., Monin, J.-L., \& Lodieu, N. 2004, \aap, 425, 927
\bibitem[n(0)]{} Knez, C., Boogert, A.C., Pontoppidan, K.M., et al. 2005,
\apj, 635, L145
\bibitem[n(0)]{} Knude, J. 2005, priv. comm.
\bibitem[n(0)]{} K\'osp\'al, \'A., \'Abrah\'am, P., Prusti, T., et al. 2007, \aap, 470,
211
\bibitem[n(0)]{} Lahuis, F., van Dishoeck, E.F., Blake, G. A., et al. 2007,
~\apj, 665, 492
\bibitem[n(0)]{} Larsson, B., Liseau, R., \& Men'shchikov, A.B. 2002,  \aap, 386,
1955
\bibitem[n(0)]{} Larsson, B., Liseau, R., Men'shchikov, A.B., et al. 2000, \aap, 363, 253
\bibitem[n(0)]{} Little, L.J., Brown, A.T., McDonalds, G.H., Riley, P.W., \&
Matheson, D.N. 1980, \mnras, 193, 115
\bibitem[n(0)]{} Lodieu, N., Caux, E., Monin, J.-L., \& Klotz, A. 2002, \aap, 383, L15
\bibitem[n(0)]{} Loren, R.B., Evans, N.J., \& Knapp, G.R. 1979, \apj, 234, 932
\bibitem[n(0)]{} Loren, R.B., Plambeck, R.L., Davis, J.H., \&  Snell, R.L. 1981,
\apj, 245, 495
\bibitem[n(0)]{} Loren, R.B. \& Wootten, A. 1980, \apj, 242, 568
\bibitem[n(0)]{} Mardones, D., Myers, P.C., Tafalla, M., et al. 1997, \apj, 489,
719
\bibitem[n(0)]{} McMullin, J.P., Mundy, L.G., Blake, G.A., et al. 2000, \apj, 536,
845
\bibitem[n(0)]{} McMullin, J.P., Mundy, L.G., Wilking, B.A., Hezel, T., \& Blake, G.A.
1994, \apj, 424, 222
\bibitem[n(0)]{} Mirabel, I.F., Ruiz, A., Rodr\'\i guez L.F., \& Cant\'o, J. 1987,
\apj, 318, 729
\bibitem[n(0)]{} Moscadelli, L., Testi, L., Furuya, R.S., et al. 2006, \aap, 446, 985
\bibitem[n(0)]{} Narayanan, G., Moriarty-Schieven, G., Walker, C.K., \& Butner, H. M.
2002, \apj , 565, 319
\bibitem[n(0)]{} Nordh, H.L., van Duinen, R.J., Sargent, A.I., et al. 1982, \aap, 115, 308
\bibitem[n(0)]{} Olmi, L. \& Testi, L. 2002, \aap, 392, 1052
\bibitem[n(0)]{} \"Oberg, K.I., Boogert, A.C.A., Pontoppidan, K.M., et al. 2008
\aj, 678, 1032
\bibitem[n(0)]{} Palla, F. \& Giovanardi, C. 1989, \aap,  223, 267
\bibitem[n(0)]{} Persi, P., Palagi, F., \& Felli, M. 1994, \aap, 291, 577
\bibitem[n(0)]{} Pontoppidan, K. M., Boogert, A.C.A., Fraser, H.J., et al. 2008,
\aj, 678, 1005
\bibitem[n(0)]{} Pontoppidan, K. M., Dartois, E.. van Dishoeck, E. F,
Thi, W.-F., \& d'Hendecourt, L. 2003a, \aap, 404, L17
\bibitem[n(0)]{} Pontoppidan, K. M., Dullemond, C.P., Blake, G.A., et al. 2007b,
\apj, 656, 980
\bibitem[n(0)]{} Pontoppidan, K. M., Dullemond, C.P., Blake, G.A., et al. 2007c,
\apj, 656, 991
\bibitem[n(0)]{} Pontoppidan, K. M., Fraser, H.J., Dartois, E. et al. 2003b,
\aap, 408, 981
\bibitem[n(0)]{} Pontoppidan, K. M., van Dishoeck, E. F., \& Dartois, E. 2004, \aap, 426, 925
\bibitem[n(0)]{} Preibisch, T. 1998, \aap, 338, L25
\bibitem[n(0)]{} Preibisch, T. 1999, \aap, 345, 583
\bibitem[n(0)]{} Preibisch, T. 2003, \aap, 410, 951
\bibitem[n(0)]{} Preibisch, T. 2004, \aap, 428, 569
\bibitem[n(0)]{} Racine, R. 1968, \aj, 73, 233
\bibitem[n(0)]{} Reipurth, B. \& Eiroa, C. 1992, \aap, 256, L1
\bibitem[n(0)]{} Reipurth, B., Nyman L.-\AA, \& Chini, R. 1996, \aap, 314, 258
\bibitem[n(0)]{} Reipurth, B., Rodr\'\i guez, L.F., Anglada, G., \& Bally, J. 2004,
\aj, 127, 1736
\bibitem[n(0)]{} Ripepi, V., Bernabei, S., Marconi, M., et al. 2007 \aap , 462, 1023
\bibitem[n(0)]{} Rodr\'\i guez, L.F., Curiel, S., Moran, J.M., et al. 1989, \apj, 346, L85
\bibitem[n(0)]{} Rodr\'\i guez, L.F., Moran, J.M., Dickinson, D.F., \& Gyulbudaghian,
A.L. 1978 \apj, 226, 115
\bibitem[n(0)]{} Rodr\'\i guez, L.F., Moran, J.M., Ho. P.T.P., \& Gottlieb, E.W. 1980,
\apj, 235, 845
\bibitem[n(0)]{} Rostopchina, A.N., Grinin, V.P., \& Shakhovskoi, D.N. 2001, Astron. Rep.
45, 60
\bibitem[n(0)]{} Schmeja, S., Klessen, R.S., \& Froebich, D. 2005, \aap, 437, 911
\bibitem[n(0)]{} Schnee, S.L., Ridge, N.A., Goodman, A. A., \& Li,
J.G. 2005 \apj, 634, 442
\bibitem[n(0)]{} Sharpless, S. 1959, \apjs, 4, 257
\bibitem[n(0)]{} Siebenmorgen, R. \& Kr\"ugel, E. 2000, \aap, 364, 625
\bibitem[n(0)]{} Smith K., Guedel M., \& Benz A.O. 1999, \aap, 349, 475
\bibitem[n(0)]{} Snell, R.L. \& Bally, J. 1986, \apj, 303, 683
\bibitem[n(0)]{} Sogawa, H., Tamura, M., Gatley, I., \& Merrill, K.M. 1997, \aj, 113,
1057
\bibitem[n(0)]{} Strai\u{z}ys, V., \u{C}ernis, K., \& Barta\u{s}i\={u}t\.{e}, S.,
1996, Baltic Astronomy, 5, 125
\bibitem[n(0)]{} Strai\u{z}ys, V., \u{C}ernis, K., \& Barta\u{s}i\={u}t\.{e}, S.,
2003, \aap, 405, 585
\bibitem[n(0)]{} Strom, S. E., Grasdalen, G.L., \& Strom, K.M. 1974, \apj, 191, 111
\bibitem[n(0)]{} Strom, S. E., Vrba, F.J., \& Strom, K.M. 1976, \aj, 81, 638
\bibitem[n(0)]{} Takano, T. 1986, \apj, 303, 349
\bibitem[n(0)]{} Testi, L. \& Sargent, A.I. 1998, \apj, 508, L91
\bibitem[n(0)]{} Testi, L., Sargent, A.I., Olmi, L., \& Onello, J.S. 2000, \apj, 540,
L53
\bibitem[n(0)]{} Torrelles, J.M., Gomez, J.F., Curiel, S., et al.  1992, \apj, 384, 59
\bibitem[n(0)]{} Torrelles, J.M., Ho, P.T.P., Rodr\'\i guez, L.F., Canto, J.,
\& Verdes-Montenegro, L. 1989, \apj, 346, 756
\bibitem[n(0)]{} Ungerechts, H. \& G\"usten, R. 1984, \aap, 131, 177
\bibitem[n(0)]{} van den Bergh, S. 1966, \aj, 71, 990
\bibitem[n(0)]{} Ward-Thompson D. \&  Buckley H.D. 2000, \mnras , 327, 95
\bibitem[n(0)]{} Warren-Smith, R.F., Draper, P.W., \& Scarrott, S.M. 1987,
\mnras, 227, 749
\bibitem[n(0)]{} White, G.J., Casali, M.M., \& Eiroa, C. 1995, \aap, 298, 594
\bibitem[n(0)]{} Williams, J.P. \& Myers, P.C. 1999, \apj, 518, L37
\bibitem[n(0)]{} Williams, J.P. \& Myers, P.C. 2000, \apj, 537, 891
\bibitem[n(0)]{} Winston, E., Megeath, S.T.,\&  Wolk, S.J., et al. 2007,
\apj, 669, 493
\bibitem[n(0)]{} Wolf-Chase, G.A., Barsony, M., Wootten H.A., et al. 1998, \apj, 501, L193
\bibitem[n(0)]{} Worden, S.P. \& Grasdalen, G.L. 1974, \aap, 34, 37
\bibitem[n(0)]{} Wu,  J., Dunham, M.M., Evans, N.J., II, Bourke, T.L., \& Young,
C.H. 2007, \aj, 133, 1560
\bibitem[n(0)]{} Wu, J.W., Wu, Y.F., Wang, J.Z., \& Cai, K. 2002, Chin. J. Astron.
Astrophys. 2, 33
\bibitem[n(0)]{} Zhang, C.Y. Laureijs, R.J., \& Clark, F.O. 1988a, \aap, 196, 236
\bibitem[n(0)]{} Zhang, C.Y. Laureijs, R.J., Clark, F.O., \& Wesselius, P.R. 1988b,
\aap, 199, 170
\bibitem[n(0)]{} Ziener, R. \& Eisl\"offel, J. 1999, \aap, 347, 565
\end{thebibliography}
\end{document}